\documentclass[twocolumn,showpacs,prc]{revtex4}       

\newcommand{ \be }{\begin{equation}}    
\newcommand{ \ee }{\end{equation}}    
\newcommand{ \bea }{\begin{eqnarray}}    
\newcommand{ \eea }{\end{eqnarray}}    
\newcommand{ \la }{\langle}    
\newcommand{ \ra }{\rangle}    
\newcommand{ \eps }{{\varepsilon}}

\usepackage{graphicx}
\usepackage{dcolumn}
\usepackage{bm}
       
\begin{document}       
\voffset=0.5 in    
       
       
\title{    
\begin{flushright}    
{{\small \sl version 27 \\ \today \\ for Phys. Rev. C }}    
\end{flushright}    
Elliptic flow from two- and four-particle correlations in Au + Au     
collisions at $\sqrt{s_{NN}} = 130$ GeV     
}    
       
\author{
C.~Adler$^{11}$, Z.~Ahammed$^{23}$, C.~Allgower$^{12}$, J.~Amonett$^{14}$,
B.D.~Anderson$^{14}$, M.~Anderson$^5$, G.S.~Averichev$^{9}$, 
J.~Balewski$^{12}$, O.~Barannikova$^{9,23}$, L.S.~Barnby$^{14}$
% ...
\begin{center}(STAR Collaboration)\end{center}
}

\affiliation{
% ...
$^5$University of California, Davis, California 95616 \\
% ...
$^{9}$Laboratory for High Energy (JINR), Dubna, Russia \\
% ...
$^{11}$University of Frankfurt, Frankfurt, Germany \\
$^{12}$Indiana University, Bloomington, Indiana 47408 \\
% ...
$^{14}$Kent State University, Kent, Ohio 44242 \\
% ...
$^{23}$Purdue University, West Lafayette, Indiana 47907 \\
% ...
}

\date{\today}       
    
\begin{abstract}       
Elliptic flow holds much promise for studying the early-time
thermalization attained in ultrarelativistic nuclear collisions. 
Flow measurements also provide a means of distinguishing between 
hydrodynamic models and calculations which approach the low density 
(dilute gas) limit. Among the effects that can complicate the
interpretation of elliptic flow measurements are azimuthal
correlations that are unrelated to the reaction plane (non-flow
correlations). Using data for Au + Au collisions at $\sqrt{s_{NN}} = 130$ GeV 
from the STAR TPC, it is found that four-particle correlation analyses
can reliably separate flow and non-flow correlation signals. The
latter account for on average about 15\% of the observed
second-harmonic azimuthal correlation, with the largest relative
contribution for the most peripheral and the most central collisions.
The results are also corrected for the effect of flow variations
within centrality bins.  This effect is negligible for all but the
most central bin, where the correction to the elliptic flow is about a
factor of two.  A simple new method for two-particle flow analysis
based on scalar products is described. An analysis based on the
distribution of the magnitude of the flow vector is also described.
\end{abstract}       
       
    
\maketitle       
       
\section{\label{sec:level1}INTRODUCTION}       
       
In non-central heavy-ion collisions, the initial spatial deformation
due to geometry and the pressure developed early in the collision
causes azimuthal momentum-space anisotropy, which is correlated with
the reaction plane \cite{Reis97, Herr99, OlliQM98, Palaiseau}.
Measurements of this correlation, known as anisotropic transverse
flow, provide insight into the evolution of the early stage of a
relativistic heavy-ion collision~\cite{Sorge97}.  Elliptic flow is
characterized by the second harmonic coefficient $v_2$ of an azimuthal
Fourier decomposition of the momentum distribution
\cite{Olli92, Volo96, Posk98}, and has been observed and extensively    
studied in nuclear collisions from sub-relativistic energies on up to
RHIC. At top AGS and SPS energies, elliptic flow is inferred to be a
relative enhancement of emission {\it in} the plane of the reaction.
Elliptic flow is developed mostly in the first several fm$/c$ (of the
order of the size of nuclei) after the collision and thus provides
information about the early-time thermalization achieved in the
collisions \cite{STAR01}.  Generally speaking, large values of flow
are considered signatures of hydrodynamic behavior~\cite{Olli92,Teaney02,Teaney01}
although an alternative approach~\cite{ZiWei02,Ko02, Molnar,Zabrodin, Huma02} 
is also argued to be consistent with the large elliptic flow for pions 
and protons at RHIC~\cite{STAR01b}.  Models in which the colliding nuclei 
resemble interacting volumes of dilute gas --- the low density
limit~\cite{Heiselberg} (LDL) --- represent the limit of mean free
path that is the opposite of hydrodynamics.  It remains unclear to
what extent the LDL picture can describe the data at RHIC, and
valuable insights can be gained from mapping out the conditions under
which hydrodynamic and LDL calculations can reproduce the measured
elliptic flow.
 
Anisotropic flow refers to correlations in particle emission with
respect to the reaction plane.  The reaction plane orientation is not
known in experiment, and anisotropic flow is usually reconstructed
from the two-particle azimuthal correlations.  But there are several
possible sources of azimuthal correlations that are unrelated to the
reaction plane --- examples include correlations caused by resonance
decays, (mini)jets, strings, quantum statistics effects, final state
interactions (particularly Coulomb effects), momentum conservation,
etc.  The present study does not distinguish between the various
effects in this overall category, but classifies their combined effect
as ``non-flow'' correlations.
       
Conventional flow analyses are equivalent to averaging over correlation     
observables constructed from pairs of particles.  When such analyses     
are applied to relativistic nuclear collisions where particle    
multiplicities can be as high as a few thousand, the possible new    
information contained in multiplets higher than pairs remains untapped.      
A previous study of high-order flow effects focused on measuring the    
extent to which all fragments contribute to the observed flow signal    
\cite{Jian92}, and amounted to an indirect means of separating flow    
and non-flow correlations.  Given that flow analyses based on pair
correlations are sensitive to both flow and non-flow effects, the
present work investigates correlation observables constructed from
particle quadruplets.  The cumulant formalism removes the lower-order
correlations which are present among any set of four particles,
leaving only the effect from the so-called ``pure'' quadruplet
correlation.  The simplest cumulant approach, in terms of both concept
and implementation, partitions observed events into four subevents.
In the present study, the four-subevent approach is demonstrated, but
our main focus is on a more elaborate cumulant method, developed by
Borghini, Dinh and Ollitrault \cite{Olli00, Olli01}. 
There are indications that non-flow effects     
contribute at a negligible level to the four-particle cumulant     
correlation\cite{Olli00, Olli01}, making it unnecessary to continue to even 
higher orders for the purpose of separating the flow and non-flow signals.    
This observation is confirmed by our Monte-Carlo simulations
       
In this analysis the observed multiplicity of charged particles within
the detector acceptance is used to characterize centrality.  This leads to 
some fluctuations of the impact parameter and, correspondingly, of the
elliptic flow within each centrality bin, especially in the bin of
highest multiplicity.  In the present study, a correction is applied
to reduce a possible bias in the measurements of the mean elliptic
flow due to impact parameter fluctuations in the centrality bins to an
insignificant level.

The present study begins with a review of the standard pair
correlation method, and provides details concerning the approach
adopted in earlier STAR publications \cite{STAR01, STAR01b} for
treating non-flow correlations.  A new method of pair flow analysis
using the scalar product of flow vectors also is introduced.  In the
conventional method, a flow coefficient is calculated by the mean
cosine of the difference in angle of two flow vectors.  In the scalar
product method, this quantity is weighted by the lengths of the
vectors.  The new method offers advantages, and is also simple to
apply. Also, an analysis in terms of the distribution of the magnitude
of the flow vector is discussed.
       
\begin{figure}[ht]      
\resizebox{20pc}{!}{\includegraphics{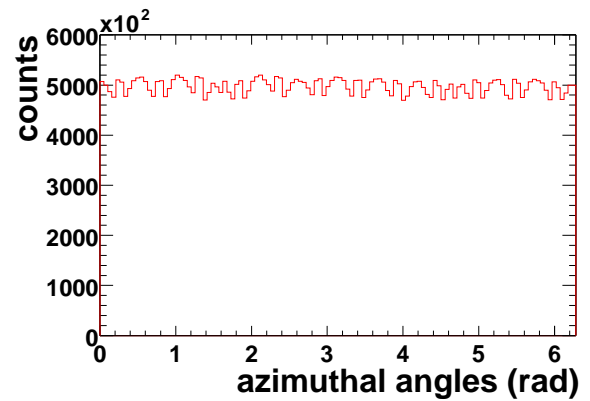}}      
\caption{\label{fig:PhiDist} The azimuthal angle distribution 
of tracks from minimum bias events. Dips are due to the
reduced efficiency at sector boundaries of STAR TPC.
}
\end{figure} 

Measurements presented in this paper are based on Au+Au data at    
$\sqrt{s_{NN}}$ = 130 GeV recorded by STAR (Solenoidal Tracker At    
RHIC) during the summer of 2000.  A detailed description of the    
detector in its year-one configuration can be found elsewhere    
\cite{STAR99}. The main feature of the STAR TPC relevant to this 
analysis is its full azimuthal coverage (see Fig.~\ref{fig:PhiDist}). 
The analysis is based on 170000 events    
corresponding to a minimum bias trigger.  Events with a primary vertex
beyond 1 cm radially from the beam or 75 cm longitudinally from the
center of the Time Projection Chamber (TPC) were excluded.  Within the
selected events, tracks were used for the estimation of the flow
vector if all five of the following conditions were satisfied: they
passed within 2 cm of the primary vertex, they had at least 15 space
points in the TPC, the ratio of the number of space points to the
expected maximum number of space points was greater than 0.52,
pseudorapidity $|\eta| < 1.3$, and transverse momentum $0.1 < p_t <
2.0$ GeV$/c$. Particles over a wider range in $\eta$ and $p_t$ were
correlated with this flow vector as shown in the graphs
below. Centrality is characterized in eight bins of charged particle
multiplicity, $n_{\rm ch}$, divided by the maximum observed charged
multiplicity, $n_{\rm max}$, with a more stringent cut $|\eta| < 0.75$
imposed only for this centrality determination.  The above cuts are
essentially the same as used in the previous STAR studies of elliptic
flow \cite{STAR01, STAR01b}.

\section{\label{sec:twoPart} Two-particle correlation methods}       
       
Anisotropic transverse flow manifests itself in the distribution of
$\phi' = \phi - \Psi$, where $\phi$ is the measured azimuth for a
track in detector coordinates, and $\Psi$ is the azimuth of the
estimated reaction plane in that event. The observed anisotropies are
described by a Fourier expansion,
\begin{equation}       
dN/d\phi' \propto 1 + 2v_{1,obs} \cos \phi' + 2v_{2,obs} \cos 2\phi' + ... \,.       
\end{equation}       
Each measurable harmonic can yield an independent estimate $\Psi_n$ of
the event reaction plane via the event flow vector $Q_n$:
\bea      
Q_n \cos n\Psi_n &=& \sum_i \cos n\phi_i, \nonumber    
\\        
Q_n \sin n\Psi_n & =& \sum_i \sin n\phi_i \,,         
\label{eq:Q_n}       
\eea       
where the sums extend over all particles in a given event. The observed 
values of $v_{n,obs}$ corrected for the reaction plane resolution yield 
$v_n$\cite{Posk98}. Below we will also use the representation of the 
flow vector as a complex number with real and imaginary parts equal to 
$x$ and $y$ components defined in Eq.~(\ref{eq:Q_n}):
\be   
Q_n= \sum_i u_{n,i},   
\label{eq:Q_n_2}
\ee   
where $u_{n,i}=e^{in\phi_i}$ is a unit vector associated with the
$i$-th particle; its complex conjugate is denoted by $u_{n,i}^*$.

\subsection{Correlation between flow angles from different subevents.     
Estimate of non-flow effects.}    
       
In order to report anisotropic flow measurements in a
detector-independent form, it is customary to divide each event into
two subevents and determine the resolution of the event plane by
correlating the $Q_n$ vector for the subevents \cite{Dani85, Posk98}.
In order to estimate the contribution from different non-flow effects
one can use different ways of partitioning the entire event into two
subevents.  The partition according to particle charge should be more
affected by resonance decay effects because the decay products of
neutral resonances have opposite charge.  The partition using two
(pseudo)rapidity regions (better separated by $\Delta y \ge 0.1$)
should greatly suppress the contribution from quantum statistics
effects and Coulomb (final state) interactions.
   
\begin{figure}[ht]       
\resizebox{20pc}{!}{\includegraphics{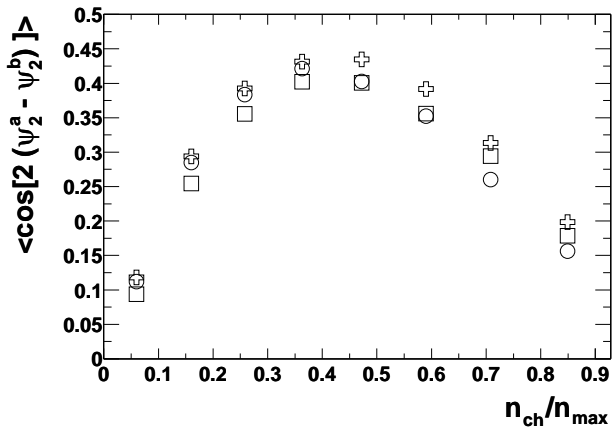}}       
\caption{\label{fig:DiffSubCosPsi} Correlation between the event     
plane angles determined from pairs of subevents partitioned randomly
(circles), partitioned with opposite sign of pseudorapidity (squares)
and partitioned with opposite sign of charge (crosses).  The
correlation is plotted as a function of centrality, namely charged
particle multiplicity $n_{\rm ch}$ divided by the maximum observed
charged multiplicity, $n_{\rm max}$.  }
\end{figure}

Another important observation for the estimate of the non-flow effects
is their dependence on centrality.  The correlation between two
subevent flow angles is
\bea   
\la \cos(2(\Psi_2^{(a)}-\Psi_2^{(b)})) \ra & \approx    
& \la \frac{ \sum_{i=1}^{M_{sub}} u_i } { \sqrt{M_{sub} }} \cdot    
\frac{ \sum_{j=1}^{M_{sub}} u_j^* } { \sqrt{M_{sub} }} \ra  \nonumber \\   
& = & \frac{ M_{sub}    
M_{sub} }    
{ M_{sub} }  \la u_i u_j^* \ra   \nonumber \\    
& \propto & M_{sub}(v_2^2 + \delta_2),   
\label{eq4}
\eea       
where $M_{sub}$ is the multiplicity of a sub-event, and $\delta_2$
denotes the non-flow contribution to two-particle correlations.  For
correlations due to small clusters, which are believed responsible for
the dominant non-flow correlations~\cite{Olli00}, the strength of the
correlation should scale in inverse proportion to the total
multiplicity.  Since the subevent multiplicity is proportional to the
total multiplicity, we can define $\tilde{\delta_2}$ to be the
multiplicity independent non-flow effect: $\delta_2 =\tilde{\delta_2}
/M_{\rm sub}$.  Collecting terms, we arrive at
\be    
\la \cos(2(\Psi_2^{(a)}-\Psi_2^{(b)})) \ra \propto     
M_{\rm sub} v_2^2 + \tilde{\delta_2}.    
\label{egtilde}    
\ee       
What is important is that the non-flow contribution to $\la    
\cos(2(\Psi_2^{(a)}-\Psi_2^{(b)})) \ra $ is approximately independent    
of centrality.  The typical shape of $ \la    
\cos(2(\Psi_2^{(a)}-\Psi_2^{(b)})) \ra $ for flow (see, for example,    
Fig.~\ref{fig:DiffSubCosPsi}) is peaked at mid-central events due to
the fact that for peripheral collisions, $M_{\rm sub}$ is small, and
for central events, $v_2$ is small.  In the previous
estimates~\cite{STAR01,STAR01b} of the systematic errors, we have set
the quantity $\tilde{\delta_2}=0.05$.  The justification for this
value was the observation of similar correlations for the first and
higher harmonics (we have investigated up to the sixth harmonic).  One
could expect the non-flow contribution to be of similar order of
magnitude for all these harmonics, and HIJING~\cite{HIJING}
simulations support this conclusion.  Given the value
$\tilde{\delta_2} = 0.05$, one simply estimates the contribution from
non-flow effects to the measurement of $v_2$ from the plot of $\la
\cos(2(\Psi_2^{(a)}-\Psi_2^{(b)})) \ra $ using Eq.~(\ref{egtilde}).    

Figure~\ref{fig:DiffSubCosPsi} shows the event plane correlation between
two subevents, for each of three different subevent partitions.  In
central events, it is seen that the correlation is stronger in the
case of subevents with opposite sign of charge compared to subevents
partitioned randomly.  This pattern might be due to resonance decays
to two particles with opposite charge.  The spread of the results for
different subevent partitions is about 0.05, which is in accord with
the number used for the estimates of the systematic errors.
   
\begin{figure}[ht]       
\resizebox{20pc}{!}{\includegraphics{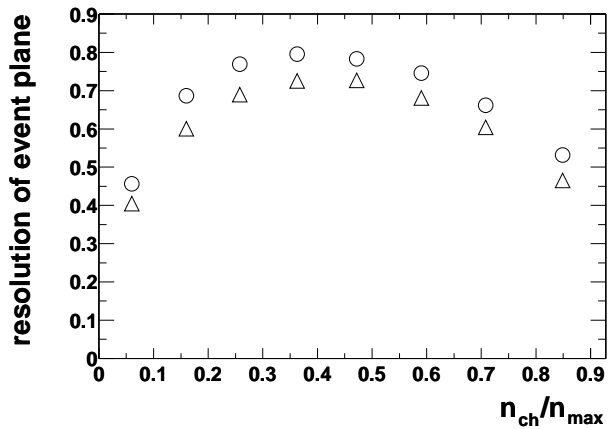}}       
\caption{\label{fig:Reso_Std}     
The event plane resolution for full events as a function of
centrality, using randomly partitioned subevents with (circles) and
without (triangles) $p_t$ weight.}
\end{figure}

\begin{figure}[ht]       
\resizebox{20pc}{!}{\includegraphics{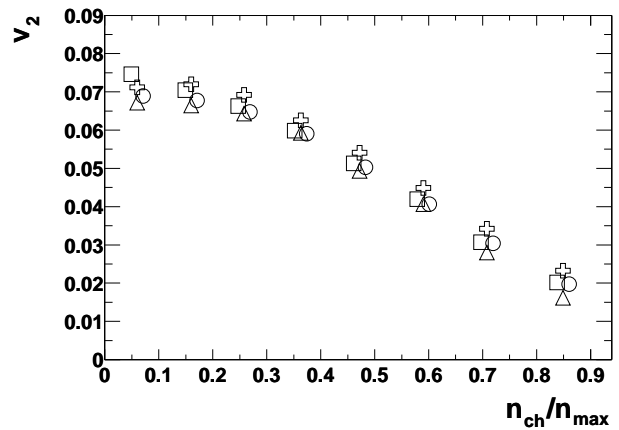}}       
\caption{\label{fig:v2cent_Std}     
Elliptic flow signal $v_2$ as a function of centrality, from study of
the correlation between particle pairs consisting of randomly chosen
particles (circles), particles with opposite sign of charge (crosses),
particles with the same sign of charge (triangles) and particles with
opposite sign of pseudorapidity (squares).  }
\end{figure}

The event plane resolution for full events is defined as $\langle \cos
( n (\Psi_{measure} - \Psi_{true}) ) \rangle$, in which
$\Psi_{measure}$ and $\Psi_{true}$ are azimuthal angles for the
measured reaction plane and the ``true'' reaction plane,
respectively. The resolution with $p_t$ weighting (see
Section~\ref{weighting}) can reach as high as 0.8, as shown in
Fig.~\ref{fig:Reso_Std}.  The $v_2$ as a function of centrality is
shown in Fig.~\ref{fig:v2cent_Std}, using different prescriptions to
partition the particles into subevents.  Again, partitioning into
subevents with opposite sign of charge yields the highest elliptic
flow signal, presumably because of neutral resonance ($\rho^0$, etc.)
decay.

\subsection{\label{weighting}Weighting}       

If Eq. (\ref{eq:Q_n_2}) is generalized to the form $Q_n = \sum_i w_i
u_i$, where the $w_i$ are weights adjusted to optimize the event plane
resolution~\cite{Dani95, Posk98}, then $u_i$ should be replaced by
$w_i u_i$ for all equations in this paper, and $M$ should be replaced
by $\sum_i w_i^2$ throughout Sec.~\ref{sec:twoPart}, and
Sec.~\ref{sec:fourSubCumul}.

\begin{figure}[ht]       
\resizebox{20pc}{!}{\includegraphics{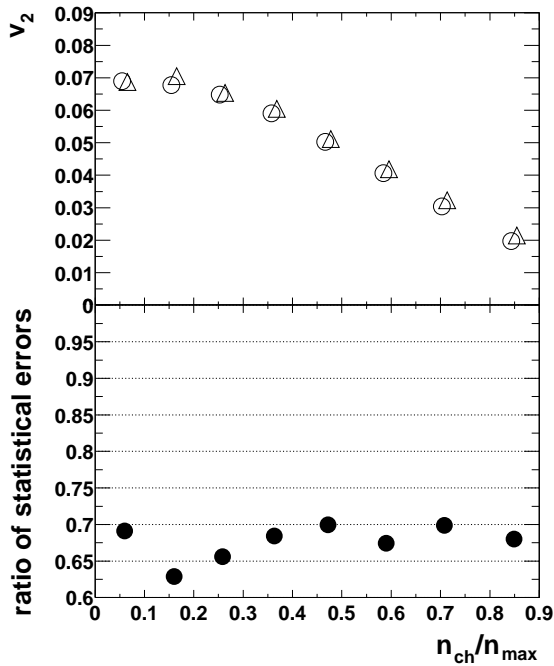}}       
\caption{\label{fig:PtWgtUnitWgt_Std} The upper panel shows $v_2$ versus
centrality using the conventional method, where the circles and
triangles represent $v_2$ with and without $p_t$ weighting,
respectively.  The statistical error is smaller than the symbol size.
The lower panel shows the statistical error on $v_2$ with $p_t$
weighting divided by the same without weighting. }
\end{figure}

The best weight $w_i(\eta, p_t)$ is $v_2(\eta, p_t)$
itself~\cite{Olli00}.  In practice, since we know that $v_2$ is
approximately proportional to $p_t$ up to about 2 GeV$/c$, it is
convenient to use $p_t$ as the weight.  It is found that $p_t$
weighting can reduce the statistical error significantly, as
demonstrated in Fig.~\ref{fig:PtWgtUnitWgt_Std}.
       
\subsection{\label{sec:Scalar}Scalar product flow analysis}       
       
In a new scalar product method \cite{ArtSergeiLBLReport}, each event
is partitioned into two subevents, labeled by the superscripts $a$ and
$b$.  The correlation between two subevents is
\begin{equation}       
\langle  Q^a_n   {Q^b_n}^* \rangle     
= \langle v^2_n M^a M^b \rangle \,  ,        
\label{eq5}   
\end{equation}       
where $M^a$ and $M^b$ are the multiplicities for subevents $a$ and
$b$, respectively.  The vectors $ Q^a_n$ and $ Q^b_n$ are constructed
for the appropriate subevent as per Eq.~(\ref{eq:Q_n}).
       
\begin{figure}[ht]       
\resizebox{20pc}{!}{\includegraphics{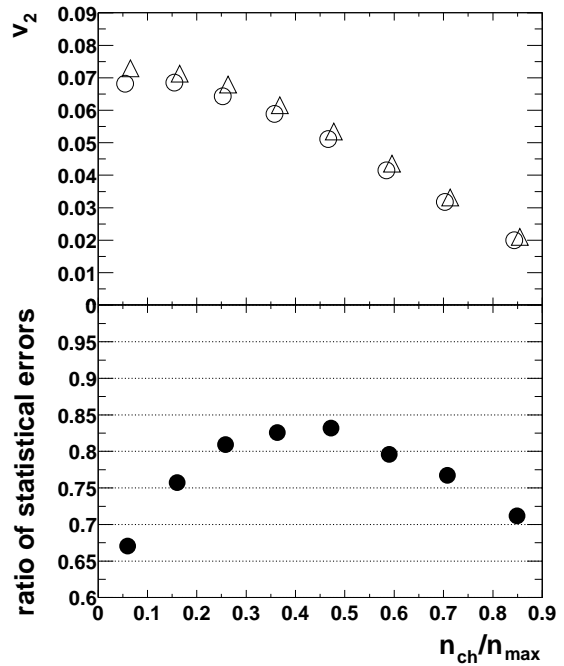}}       
\caption{\label{fig:v2scalar} The upper panel presents $v_2$ versus     
centrality from the scalar product method (triangles) and the
conventional random subevent method (circles).  All statistical errors
are smaller than the symbol size.  The statistical error for the
scalar product method divided by that for the conventional method is
shown in the lower panel. }
\end{figure}        
    
Given the above, the flow relative to the true reaction plane can be
readily calculated from unit momentum vectors $ u_{n,i}(\eta, p_t)$ of
the analyzed tracks by using Eq. (\ref{eq5}) for the particle relative
to the $2M$ other particles, and then dividing by the square root of
Eq. (\ref{eq5}) for the subevents. This gives
\begin{equation}       
v_n(\eta, p_t)     
= \frac{ \langle  Q_n   u_{n,i}^*(\eta, p_t) \rangle }       
  {2 \sqrt{ \langle Q^a_n  {Q^b_n}^* \rangle } } \,.       
\end{equation}       
Auto-correlations are removed by subtracting particle $i$ in the
calculation of $Q_n$ when taking the scalar product with $u_{n,i}$.
This method weights events with the magnitude of the $Q_n$
vector, and if $Q_n$ is replaced by its unit vector, the above reduces
to $\langle \cos n(\phi-\Psi) \rangle$, the conventional correlation
method.
     
Figure~\ref{fig:v2scalar} demonstrates that the results from the scalar
product method are indeed very close to the ones of the conventional
method. In this calculation, the subevents are generated by random
partitioning. However, the detailed comparison of two results reveals
a small systematic difference. The difference might have origin in the
approximations (the Central Limit Theorem) used in the conventional
method and that are not required in the scalar product method. In
addition, the scalar product method has the benefit of smaller
statistical errors and is very simple to implement.

\section{Distribution in the magnitude of the flow vector}

In this section we study elliptic flow by analysis of the distribution
in the magnitude of the flow vector. The method was used by the E877 
Collaboration at the AGS for the first
observation of anisotropic flow at ultra-relativistic nuclear
collisions~\cite{E877PRL94}.  This method is based on the observation
that anisotropic flow strongly modifies the distribution of the
magnitude of the flow vector~\cite{Volo96,Olli95,Posk98,Olli00}.  Very
strong flow leads to the distribution, $dP/(Q_n dQ_n)$ with a local
minimum at $Q=0$, which reflects the fact that for the case of strong
flow all particle momentum unit vectors are aligned in the flow
direction.  On the other hand, the non-flow effects, two and few
particle azimuthal correlations lead to an increase in the statistical
fluctuation width of the distribution. The effect can be understood by
considering the flow vector composed of many clusters but randomly
distributed in the azimuthal space.  In the limit of large
multiplicity and neglecting the contribution from higher harmonics
(for a more accurate consideration see~\cite{ArtSergeiLBLReport,Olli97,Olli00}) the
distribution can be described by~\cite{Volo96,Posk98,Olli95}:
\begin{equation}
\frac{dP}{q_n d q_n} = \frac{1}{\sigma_n^2}
       e^{\displaystyle{-\frac{v_n^2M+q_n^2}{2\sigma_n^2}}}  
       I_0(\frac{q_n v_n \sqrt{M}}{\sigma_n^2}),
\label{qdist}
\end{equation}
where $I_0$ is the modified Bessel function. We have introduced the
variable $q_n=Q_n/\sqrt{M}$, which greatly reduces the effect on the
shape of the distribution from averaging over events with different
multiplicities.  In a more general case using weights, one should use
$q_n=Q_n/(\sqrt{M \la w_i^2 \ra}$).  In this way the width of the
$q-$distribution is independent of multiplicity:
\begin{equation}
\sigma_n^2 = 0.5 (1 + g_n),
\end{equation}
with $g_n$ reflecting the change in the width of the distribution due
to non-flow effects (and to some extent to the averaging over events
with different multiplicities).

\begin{figure}[ht]       
\resizebox{20pc}{!}{\includegraphics{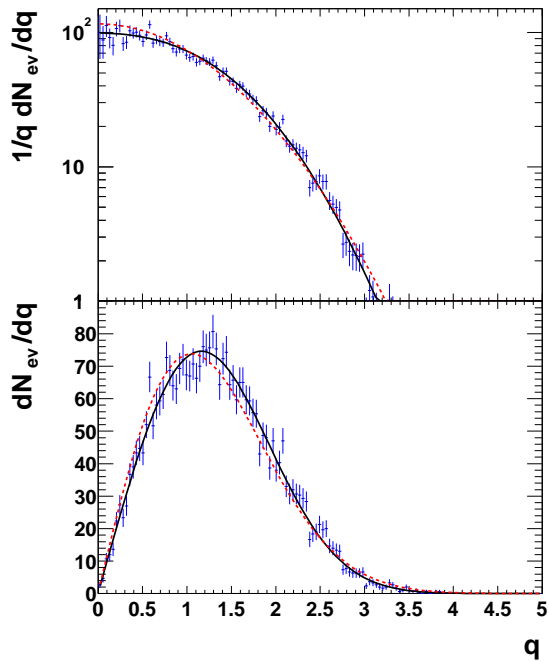}}       
\caption{\label{fig:fqdist} 
Reduced flow vector distributions for centrality bin 5 plotted in two ways. 
Solid lines correspond to the fit with two parameters,  $v_2$ and $g_2$, and
dashed lines correspond to the fit assuming zero real flow.
}
\end{figure}        

We have fitted distributions of $q_2$, the second harmonic reduced
flow vector, in two different ways.  First, the distributions in all
different centrality bins have been fitted with two independent
parameters, $v_2$ and $g_2$. The non-flow contribution parameter,
$g_2$, has been found to be in the range of 0.18 --0.32 for all
centralities except the most peripheral one.  One should not expect a
good fit for the most peripheral bin, for it is a mixture of events in
a wide multiplicity range from 20 to 100.  Better fit results for this
bin could be achieved if the bin would be split into several sub-bins
with smaller relative multiplicity variations.  The relative
multiplicity variation in the other bins is much smaller.  The $q$
distribution for the centrality bin 5 is presented in
Fig.~\ref{fig:fqdist}. The two fit functions correspond to the case of a
fit with two parameters, $v_2$ and $g_2$, and to the case of a one
parameter fit of $g_2$ for $v_2=0$. Note that the dashed curves are
systematically higher or lower than the data points in different $q$
regions. In the lower part of Fig.~\ref{fig:fqdist} one can see that the 
anisotropic flow pushes the q distribution out to larger values. If 
the flow were great enough one could select events based on the q values.

\begin{figure}[ht]       
\resizebox{20pc}{!}{\includegraphics{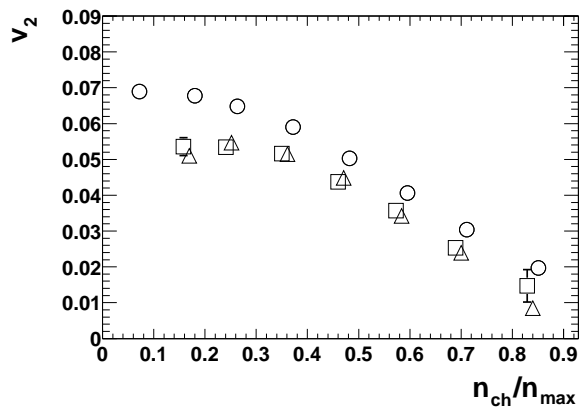}}       
\caption{\label{fig:fv2fromqdist} 
Elliptic flow as determined from the fits to the $q$ distributions in
different centrality bins. The circles are from the standard method
with random subevents.  For the squares, all the centralities were fit
separately.  For the triangles, centrality bins 2 to 8 were fit with
the same value of the non-flow parameter.  }
\end{figure} 
       
In the second method we fit $q$ distributions in centrality bins 2 to
8 simultaneously with different $v_2$ values for each centrality bin
but the same value of $g$. (This assumption is similar to the
assumption of $\tilde{\delta}=const$ in the previous section. See also
the discussion in~\cite{Olli95,Posk98,Olli00}).  We find $g=0.29 \pm
0.02$.  The results of the fits are presented in
Fig.~\ref{fig:fv2fromqdist}.  The deviation from the standard method
results are due to the non-flow contributions.

\section{Four-particle correlations}       
       
\subsection{Motivation for Cumulants}       
       
In experiments, it is necessary to rely on correlations between
particles to determine the event plane since the reaction plane is not
a direct observable.  The assumption underlying conventional pair
correlation analyses (including the scalar product method discussed in
section II.C above) is that non-flow correlations of the type
mentioned in section I are negligible compared to the flow, or at
most, are comparable to other systematic uncertainties.  In past
studies \cite{Posk98, Mai, borghiniNoflow}, non-flow correlations have
been discussed with specific reference to their origin, such as
momentum conservation, Bose-Einstein correlations, Coulomb effect,
jets, resonance decays, etc.  In the first two studies of elliptic
flow in STAR \cite{STAR01, STAR01b}, the non-flow effect from jets and
resonances was estimated using the approach explained in section II.A
above, and this established an upper limit on the non-flow
contribution to the reported $v_2$ signal.  This limit played a role
in determining the systematic error on the published measurements.
       
Anisotropic flow is a genuine multiparticle phenomenon, which
justifies use of the term {\sl collective flow}.  It means that if one
considers many-particle correlations instead of just two-particle
correlations, the relative contribution of non-flow effects (due to
few particle clusters) should decrease.  Considering many-particle
correlations, one has to subtract the contribution from correlations
due to lower-order multiplets.  Formally, one should use
cumulants \cite{Biya81, Libo89, Egge93, Olli01} instead of simple
correlation functions.  Let us explain this with an example for
four-particle correlations.  The correlation between two particles is
\be    
\la u_{n,1} u_{n,2}^* \ra \equiv \la e^{i n \phi_1} e^{- i n \phi_2} \ra    
=v_n^2 + \delta_n \,,    
\label{twonf}    
\ee    
where $n$ is the harmonic, and the average is taken over all pairs of
particles in a given rapidity and transverse momentum region, and over
all events in an event sample.  The $\delta_n$ represents the
contribution to the pair correlation from non-flow effects.
Correlating four particles, one gets
\be    
\la u_{n,1} u_{n,2} u_{n,3}^* u_{n,4}^* \ra    
= v_n^4 + 2 \cdot 2 \cdot v_n^2 \delta_n + 2 \delta_n^2 \,.    
\label{fournf}    
\ee    
In this expression, two factors of ``2'' in front of the middle term
correspond to the two ways of pairing (1,3)(2,4) and (1,4)(2,3) and
account for the possibility to have non-flow effects in the first pair
and flow correlations in the second pair and vice versa.  The factor
``2'' in front of the last term is due to the two ways of pairing.
The pure four-particle non-flow correlation is omitted from this
expression --- see the discussion below about the possible magnitude
of such a contribution.  What is remarkable is that if one subtracts
from the expression (\ref{fournf}) twice the square of the expression
(\ref{twonf}), one is left with only the flow contributions
\bea    
&& \la \la  u_{n,1} u_{n,2} u_{n,3}^* u_{n,4}^* \ra \ra    
\nonumber \\    
&&    
\equiv    
\la u_{n,1} u_{n,2} u_{n,3}^* u_{n,4}^* \ra    
-2 \la u_{n,1} u_{n,2}^* \ra ^2    
= -v_n^4 \,,    
\eea    
where the notation $\la \la ... \ra \ra$ is used for the {\sl
cumulant}.  The cumulant of order two is just $\la \la u_{n,1}
u_{n,2}^* \ra \ra =\la u_{n,1} u_{n,2}^* \ra$.
   
In flow analysis, one is interested not only in so-called ``global''
flow values, but also in differential flow as function of rapidity and
transverse momentum.  In a four-particle correlation approach, this
also can be done in a similar manner, now correlating a particle, for
example in a particular $p_t$ bin, with three particles from a common
``pool''.  Assuming that the particle ``b'' is the one from a
particular bin, one gets for a differential flow study
\be    
\la u_{n,b} u_{n,1}^* \ra = v_{n;b} v_n + \delta_{n;b} \,,    
\label{twonfpt}    
\ee    
where we have introduced the notation $v_{n;b}$ for the flow value
corresponding to the bin under study, and $\delta_{n;b}$ for the
corresponding non-flow contribution.  Then for the correlation with
three particles from the pool,
\bea    
&&    
\la u_{n,b} u_{n,1} u_{n,2}^* u_{n,3}^* \ra = \nonumber \\    
&&     
v_{n;b} v_n^3 + 2 \cdot v_n^2 \delta_{n;b} + 2     
\cdot v_n v_{n;b} \delta_{n} + 2 \delta_n \delta_{n;b} \,.    
\label{fournfpt}    
\eea    
In this case, in order to remove the non-flow contribution, one has to
subtract from (\ref{fournfpt}) twice the product of expressions
(\ref{twonf}) and (\ref{twonfpt}).
\bea    
&&    
\la u_{n,b} u_{n,1} u_{n,2}^* u_{n,3}^* \ra    
- 2 \la u_{n,b} u_{n,1}^* \ra \la u_{n,1} u_{n,2}^* \ra = \\    
&&    
-v_n^3 v_{n;b} \,. \nonumber    
\eea    
Assuming that the average flow value for the particles in the pool is
known, one gets the desired differential flow value for the particular
bin under study.
      
In Eq. (\ref{fournf}), we have neglected the contribution from the
pure four-particle correlations due to non-flow effects.  Let us now
estimate the upper limit for such a contribution.  Assume that {\sl
all} particles are produced via four-particle clusters.  All daughters
of the decay of such a cluster could in principle be within 1--2 units
of rapidity from each other.  Then the contribution would be
\be    
6f/M^3,    
\ee    
where $M$ is the {\sl total} multiplicity within those 1--2 units of
rapidity, and $f$ is $\la (\cos 2(\phi_1-\phi_2))^2 \ra$ averaged over
all cluster decay products.  Assuming a perfect alignment, $f=1$, and
multiplicity $M=1000$, this would give us a possible error in $v_2$
measurements of the order of
\be
\delta v \sim  (v_2^4+6/1000^3)^{1/4} - v_2 \, .
\ee
This would give only 3\% relative error on $v_2$ signal of 0.015, and
would drops very rapidly with increasing real $v_2$ signal.  This
calculation is for the case of 100\% of the particle production via
four-particle clusters and a perfect alignment of decay products.  A
more realistic scenario would give a much smaller estimate.

\subsection{\label{sec:fourSubCumul}Four-subevent method}    
    
In order to apply the four-particle correlation approach to the
analysis of real data, one should perform an average over all possible
quadruplets of particles in a given event.  Bearing in mind that the
average multiplicity in a central STAR event is well beyond a
thousand, it becomes a nontrivial task. The simplest solution to the
problem is the four-subevent method where one partitions all tracks
(for example, randomly) into four groups (subevents) and calculates a
flow vector for each of the groups,
\be    
Q_n=\sum_i u_{n,i} \,,    
\ee    
where the sum is over all particles in the group.  Using these    
subevents, the problem becomes much simpler computationally.  For    
example,     
\be    
\la u_{n,1} u_{n,2} u_{n,3}^* u_{n,4}^* \ra    
= \la Q_{n,1} Q_{n,2} Q_{n,3}^* Q_{n,4}^*     
/(M_1 M_2 M_3 M_4)  \ra \,,    
\label{fournff}    
\ee    
where $M_i$ are the corresponding subevent multiplicities.  The
cumulant calculation is straightforward:
\bea    
&&    
\la \la u_{n,1} u_{n,2} u_{n,3}^* u_{n,4}^* \ra \ra = \nonumber \\    
&&    
\la \frac{Q_{n,1} Q_{n,2} Q_{n,3}^* Q_{n,4}^* }{ M_1 M_2 M_3 M_4}  \ra     
\nonumber \\    
\;\;\;\;\;\;    
&&     
- 2 \left(\la Q_{n,1} Q_{n,2}^* /(M_1 M_2) \ra\right)^2 .    
\eea    
    
The four-subevent method is very simple, both in logic and in
implementation.  The price for these benefits is lower statistical
power, because the method does not take into account all possible
quadruplets. Some improvement could be reached by splitting the event
into more than four subevents and correlating all possible
combinations of four.  In the analysis of the STAR data we use eight
subevents. A more general cumulant formalism, based on the cumulant
generating function \cite{Mai,Olli01} offers advantages for a
four-particle analysis in the context of the present limited sample
size.

\subsection{Cumulant generating function}    
    
The cumulant and generating function approach offers a formal and
convenient way to study flow and non-flow contributions
systematically.  Following the method of Ref.~\cite{Olli01}, the
cumulant to order four is defined by
\begin{eqnarray}       
\langle\langle    
u_1 u_2 u_3^* u_4^*   
\rangle\rangle    
\equiv         
\langle    
u_1 u_2 u_3^* u_4^*   
\rangle - \nonumber \\       
\langle    
u_1 u_3^*   
\rangle \langle    
u_2 u_4^*   
\rangle - \langle    
u_1 u_4^*   
\rangle \langle    
u_2 u_3^*   
\rangle \,,        
\label{eq:one}       
\end{eqnarray}       
where, as above, the double angle bracket notation represents the
cumulant expression shown explicitly on the right-hand side. The
subscript for the harmonic order, $n$, has been dropped.  The cumulant
$\langle\langle u_1 u_2 u_3^* u_4^*
\rangle\rangle$ involves only pure four-particle    
correlations, since the two-particle only correlations among the    
quadruplets have been explicitly subtracted away.    
      
In the presence of flow, the cumulant becomes      
\begin{equation}      
\langle\langle    
u_1 u_2 u_3^* u_4^*   
\rangle\rangle       
= - v^4_n + O ( \frac{1}{M^3} + \frac{v^2_{2n}}{M^2} ) \,,      
\end{equation}      
where $M$ is the multiplicity of the events, the term of order $1/M^3$
represents the remaining four-particle non-flow effects, and the term
of order $v^2_{2n}/M^2$ is the contribution of the $2n$ higher
harmonic.  The cumulant to higher orders and the corresponding
generalization has also been determined~\cite{Olli01}.  Likewise, the
cumulant of order two reduces to the equivalent of a pair correlation
analysis of the conventional type.  Statistical uncertainties
associated with a cumulant analysis increase with increasing order
from 2 to 4.
      
The definition of the cumulant is simple, but it is tedious to
calculate the moments term-by-term on the right-hand side of
Eq.~(\ref{eq:one}).  Fortunately, the cumulant can be computed more
easily from the generating function~\cite{Olli01},
\begin{equation}      
G_n(z)  =  \prod_{j=1}^M       
\left( 1+ {z^* u_j + z u_j^* \over M} \right),      
\label{eq:newG0}      
\end{equation}    
where $z \equiv |z| e^{i\alpha}$ is an arbitrary complex number, $z^*$
denotes its complex conjugate. The generating function itself has no
direct physical meaning, but the coefficients of the expansion of
$\langle G_n \rangle$ in powers of $z, z^*$ yield the correlations of
interest:
\bea   
\la G_n \ra &=& 1+\la \frac{M-1}{M} \ra |z|^2 \la u_1 u_2^* \ra    
\\ \nonumber   
&+&    
\la \frac{(M-1)(M-2)(M-3)}{4M^3} \ra |z|^4 \la u_1 u_2 u_3^* u_4^* \ra +... .   
\label{eq:gen-corr}   
\eea   
One can use these correlations to construct the cumulants.  In the
limit of large $M$, $\la G_n \ra$ can be used to obtain the cumulant
generating function directly:
\begin{eqnarray}      
M\cdot\left(\langle G_n(z) \rangle^{1/M}-1\right) = \nonumber \\      
\sum_{k} {|z|^{2k}\over (k!)^2} \langle \langle       
u_1 ... u_k u_{k+1}^*...u_{2k}^* \rangle \rangle \,.      
\label{eq:defc2}      
\end{eqnarray}      
The left-hand side of Eq.~(\ref{eq:defc2}) is what is measured, and in
order to extract the cumulants on the right, $k$ equations of the form
of Eq.~(\ref{eq:defc2}) are needed to solve for $k$ undetermined
parameters.  This can be accomplished by repeating the process with
$k$ different values of $|z|$.  It is found that suggested magnitudes
of $|z|$ in Ref.~\cite{Olli01}, namely $r_0 \sqrt{p}$ with $r_0=1.5$
and $p=1,\cdots k$, are fairly good, since results from optimized
values~\cite{cumuPraticeGuide} of $r_0$ show almost no
difference. Results in this paper are by default calculated with
$r_0=1.5$. Since $M$ fluctuates from one event to the other, for
events within a multiplicity bin, we use the average value $\langle M
\rangle$ in Eq.~\ref{eq:defc2} instead of $M$.
   
For experimental analysis, it is sufficient to take the first three
terms in Eq.~(\ref{eq:defc2}). Once the cumulant has been computed,
extracting the integrated flow value is straightforward because, for
instance, $v^4_n = - \langle \langle u_1 u_2 u_3^* u_4^* \rangle
\rangle$.
      
When a non-unit weight is used, the integrated flow value described
above becomes $\la w \cos n \phi' \ra$, which is not exactly $v_n$ but
an approximation.  However, the differential flow can be calculated
exactly (see below) no matter what weight is used.  The integrated
flow with non-unit weight can be obtained by integrating the
differential flow.  All integrated flow results in this paper (except
for results from the four-subevent method) are obtained by integrating
over the differential flow.
   
For differential flow (flow in a bin of $\eta$ and/or $p_t$),    
Eq.~(\ref{eq:defc2}) is replaced by       
\begin{eqnarray}      
&& \frac{\langle       
u_d G_n(z) \rangle }{\langle G_n(z) \rangle} \equiv \nonumber \\    
&& \sum_{k,l} {z^{*k}z^{l}\over k!\,l!} \langle \langle       
u_d u_1 ... u_k u_{k+1}^*...u_{k+l}^*   
\rangle \rangle \,.      
\label{eq:defcm1}      
\end{eqnarray}      
where $u_d$ is the unit vector for a particle in the selected bin.
Following a similar procedure as in the case of the integrated flow,
the cumulant $ \langle \langle
u_d u_1 u_2^* u_3^*  
\rangle \rangle$ is computed, but it now contains the angle of the one    
particle of interest and three other particles from the pool.  Then
the differential flow is~\cite{Olli01}
\begin{equation}      
\label{eq:diffFlow}      
v_n = - \frac{ \langle \langle    
u_d u_1 u_2^* u_3^*  
\rangle \rangle}      
  {(-\langle \langle    
u_1 u_2 u_3^* u_4^*     
\rangle \rangle \,)^{    
3/4   
}}.      
\end{equation}        
     
\begin{figure}[ht]      
\resizebox{20pc}{!}{\includegraphics{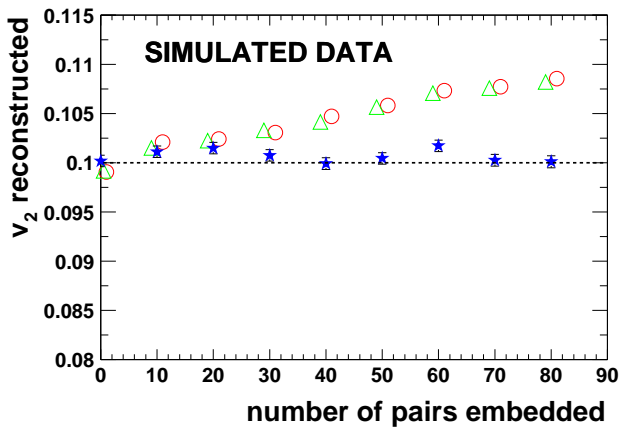}}      
\caption{\label{fig:SimV2ConstPairVary} Reconstructed $v_2$ from     
the conventional method (circles), from the 2nd-order cumulant method
(triangles), and from the 4th-order cumulant method (stars), for
simulated events as a function of number of embedded back-to-back
track pairs.  The horizontal dashed line marks the level of the true
elliptic flow $v_2 = 0.10$, as imposed on the simulated events,
including the back-to-back track pairs.  The statistical error is
smaller than the symbol size.  The multiplicity for all events is
500.}
\end{figure}       
   
Eq.~(\ref{eq:diffFlow}) is for unit weight.  It can
be easily generalized for non-unit weight, and the formula still
holds.

Some detectors have substantial asymmetry in their response as a    
function of azimuth in detector coordinates, in which case it is    
necessary to prevent distortion of the measured flow signals by 
employing one of two possible compensation methods \cite{Posk98} ---
applying a shifting transformation which recenters $Q$: $\langle Q_n
\sin n \Psi_n \rangle =0$ and $\langle Q_n \cos n \Psi_n \rangle =0$
(see Eq. (~\ref{eq:Q_n})), or applying weighting factors to force a
flat $\Psi$ distribution. In the present study, no noticeable
difference is observed with and without explicit compensation for
detector asymmetry, as expected in light of the excellent azimuthal
symmetry of the STAR TPC.  All plots 
in this paper are made without compensation for detector asymmetry. 
However, it should be noted that cumulants, as defined by the 
generating function, also correct for small anisotropies in 
the detector acceptance.  For instance, the cumulant
\begin{equation}      
\langle \langle u_1 u_2^* \rangle \rangle = \langle u_1 u_2^* \rangle
- \langle u_1 \rangle \langle u_2^* \rangle
\end{equation}        
amounts to an implementation of the shifting compensation method
mentioned above.

\subsection{Simulations}       
       
\begin{figure}[ht]      
\resizebox{20pc}{!}{\includegraphics{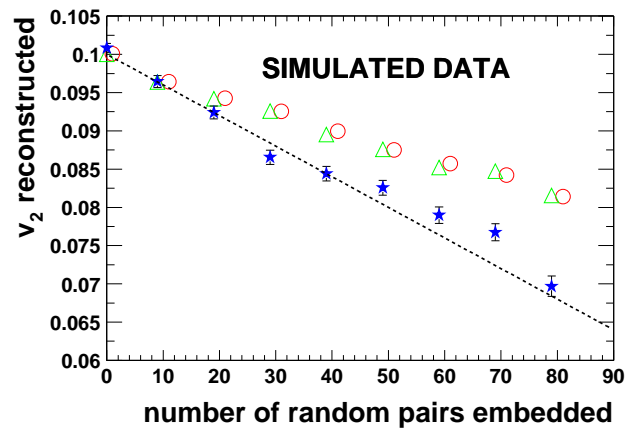}}      
\caption{\label{fig:SimRdmPairVary} Reconstructed $v_2$ from the conventional      
method (circles), from the 2nd-order cumulant method (triangles), and
from the 4th-order cumulant method (stars), for simulated events as a
function of number of embedded back-to-back track pairs.  Unlike in
the previous figure, the embedded back-to-back pairs are randomly
distributed relative to the event plane, and so the true resultant
$v_2$, indicated by the dashed line, decreases as more pairs are
embedded.  The multiplicity for all events is 500.}
\end{figure}       
      
In order to test the cumulant method as well as the analysis    
procedure, the MEVSIM~\cite{mevsim} event generator has been used to    
make events with various mixtures of flow and non-flow effects.  In    
all cases, the number of simulated events in a data set is 20k, and    
the multiplicity is 500.  Fig.~\ref{fig:SimV2ConstPairVary} shows one    
such set of simulations.  Nine data sets with $v_2 = 0.10$ were    
produced, then a simple non-flow effect consisting of embedded    
back-to-back track pairs was introduced at various levels, ranging    
from zero up to 80 pairs per simulated event.  These pairs simulate    
resonances which decay to two daughters with a large energy release.    
In Fig.~\ref{fig:SimV2ConstPairVary}, we consider the scenario where    
the embedded pairs themselves are correlated with the event plane with    
the same $v_2 = 0.10$.  Fig.~\ref{fig:SimV2ConstPairVary} shows that    
the 4th-order cumulant $v_2$ always reconstructs the expected 10\%    
$v_2$, while the $v_2$ from the pair correlation analysis methods can    
only recover the correct input if non-flow pairs are not embedded.    
     
If back-to-back pairs are instead randomly distributed in azimuth, the    
true flow should decreases and the expected variation can be computed    
knowing the number of random tracks.  Fig.~\ref{fig:SimRdmPairVary}    
shows such a simulation, and again it is found that only the 4th-order    
cumulant $v_2$ agrees with the expected elliptic flow, while the    
inferred $v_2$ based on pair correlation analyses is distorted in the    
presence of the simulated non-flow effects.  The role of resonances    
produced in real collisions may be closer to one or the other of the    
above two simulated scenarios, but in either case, the non-flow effect    
is removed by the 4th-order cumulant analysis.    
      
\begin{figure}[ht]      
\resizebox{20pc}{!}{\includegraphics{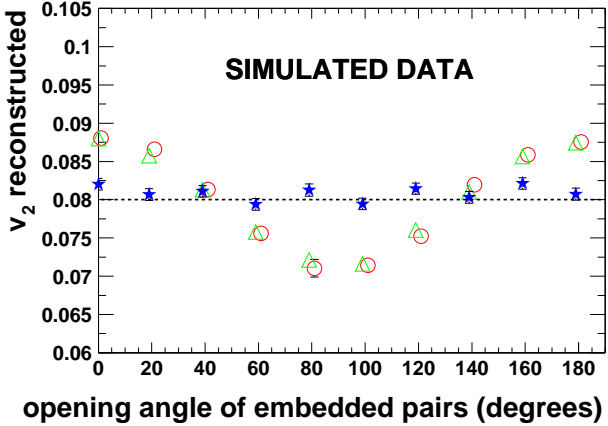}}      
\caption{\label{fig:SimAzimuAglVary} Elliptic flow from     
the conventional method (circles), from the 2nd-order cumulant method    
(triangles), and from the 4th-order cumulant method (stars), for    
simulated events as a function of azimuthal angle between the two    
tracks in each of 50 embedded pairs per event, with the 50 pairs each    
having random orientation relative to the event plane.  The horizontal    
dashed line marks the level of the true elliptic flow $v_2$ = 0.08.  }    
\end{figure}       
      
In Fig.~\ref{fig:SimAzimuAglVary}, consideration is given to the    
possible effect of resonances which decay with smaller energy release,    
having an azimuthal opening angle $\Phi$ in the laboratory.  The    
simulated events were generated with an imposed flow $v_2 = 0.08$,    
while in each event 50 pairs with the same $\Phi$ were embedded, each    
such pair having a random orientation relative to the event plane.    
Ten data sets were produced, with $\Phi$ (the abscissa in    
Fig.~\ref{fig:SimAzimuAglVary}) varying in $20^\circ$ steps between    
zero and $180^\circ$.  Again, only the 4th-order cumulant $v_2$    
(stars) recovers the true elliptic flow signal.    
      
\begin{figure}[ht]      
\resizebox{20pc}{!}{\includegraphics{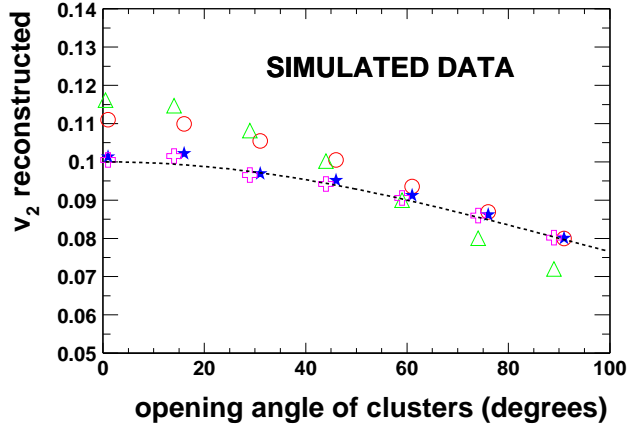}}      
\caption{\label{fig:SimV2ConstClstrOpenAglVary} Elliptic flow    
from the conventional method (circles), and from the 2nd-order    
(triangles), 4th-order (stars), and 6th-order (crosses) cumulant    
methods. This is for simulated events as a function of azimuthal angle    
between two back-to-back track pairs.  The dashed line marks the level    
of the true elliptic flow.  }    
\end{figure}       
      
In order to test how the various methods respond to non-flow    
correlations associated with four-particle clusters, the simulated    
events in Fig.~\ref{fig:SimV2ConstClstrOpenAglVary} were generated    
with an imposed flow $v_2 = 0.10$, after which 25 four-particle    
clusters were embedded in each event.  Each cluster consists of two    
back-to-back pairs with an azimuthal opening angle $\Phi$ between    
them.  Seven data sets were produced, with $\Phi$ (the abscissa in    
Fig.~\ref{fig:SimV2ConstClstrOpenAglVary}) varying in $15^\circ$ steps    
between zero and $90^\circ$.  The clusters were oriented such that a    
track bisecting $\Phi$ would contribute to the overall flow with $v_2    
= 0.10$.  The 4th-order cumulant (stars) and the 6th-order cumulant    
(crosses) both reconstruct the true elliptic flow (dotted line).  Note    
that the four-particle correlation introduced by the clusters is    
$~1/M^2$ times the pair correlation part, resulting in little    
difference between $v_2$ from the 4th- and 6th-order cumulant methods.    
This result further illustrates the point (see also the end of section     
III.A) that non-flow effects are believed to contribute at a negligible     
level to the four-particle correlation, and for this reason, there may     
be little advantage in extending cumulant analyses to orders higher     
than 4.    
      
\begin{figure}[ht]      
\resizebox{20pc}{!}{\includegraphics{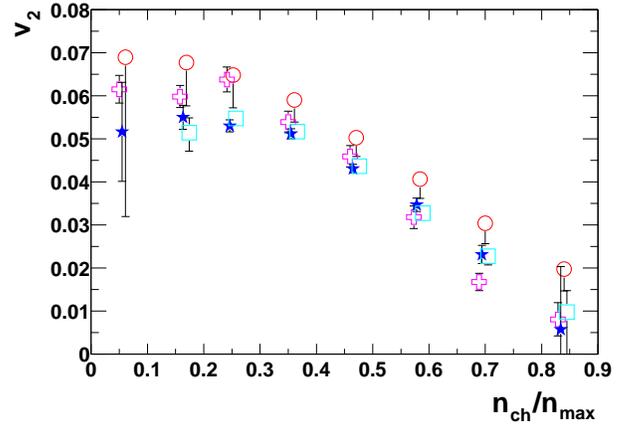}}      
\caption{\label{fig:vcent} Measured elliptic flow versus centrality     
for Au + Au at $\sqrt{s_{NN}} = 130$ GeV.  The circles show the    
conventional $v_2$ with estimated systematic uncertainty due to non-flow    
\cite{QM2001Raimond}, the stars show the 4th-order cumulant $v_2$ from    
the generating function, the crosses show the conventional $v_2$ from
quarter-events, and the squares show the 4th-order cumulant $v_2$ from
the four-subevent method.}
\end{figure}

\subsection{Results from STAR}   
       
Figure~\ref{fig:vcent} shows measured elliptic flow versus centrality,    
where the latter is characterized by charged particle    
multiplicity $n_{\rm ch}$ divided by the maximum observed charged    
particle multiplicity, $n_{\rm max}$.  The conventional $v_2$ (circles),    
the 4th-order cumulant $v_2$ from the generating function (stars), and    
the 4th-order cumulant $v_2$ from the four-subevent method (squares)    
are compared.  The cross symbols in Fig.~\ref{fig:vcent} represent the    
conventional $v_2$ signal for the case where each observed event is    
partitioned into four quarter-events, which are then analyzed like    
independent events.  All tracks in each quarter-event have the same    
sign of charge, and the same sign of pseudorapidity.  Furthermore, the    
event plane for quarter-events is constructed using only tracks with    
$p_t < 0.5$ GeV$/c$, which serves to minimize the influence of non-flow    
associated with high-$p_t$ particles.  It is clear that the non-flow    
effect is present at all centralities, and its relative magnitude is    
least at intermediate multiplicities.      
        
\begin{figure*}[htp]      
\resizebox{35pc}{!}{\includegraphics{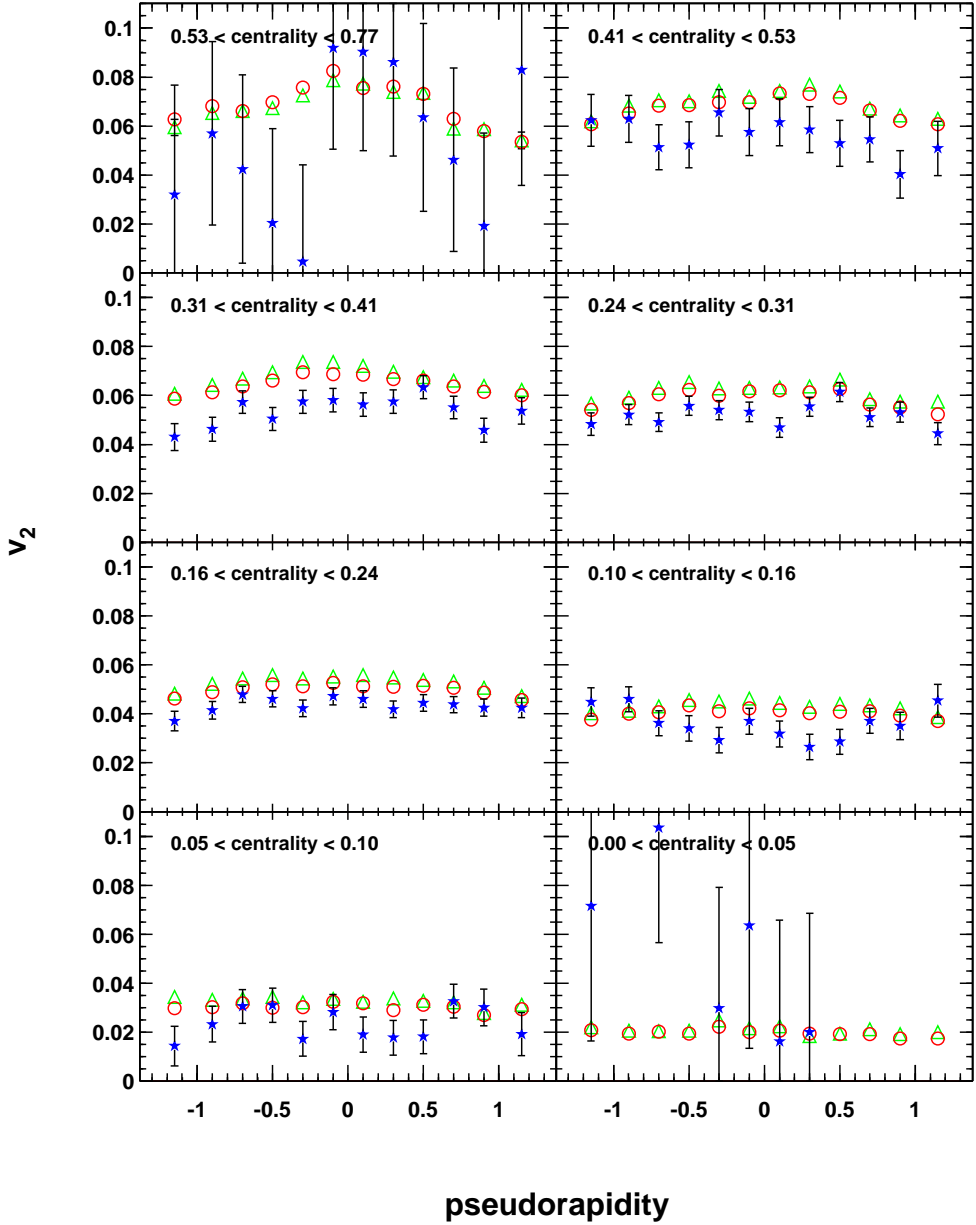}}      
\caption{\label{fig:EtaDifferentialFlowAllCent} Reconstructed $v_2$ versus    
pseudorapidity from the conventional method (circles), from the    
2nd-order cumulant method (triangles), and from the 4th-order cumulant    
method (stars), in eight centrality bins.  The upper left panel shows    
the most peripheral events, and the lower right the most central.}    
\end{figure*}       
      
\begin{figure*}[htp]      
\resizebox{35pc}{!}{\includegraphics{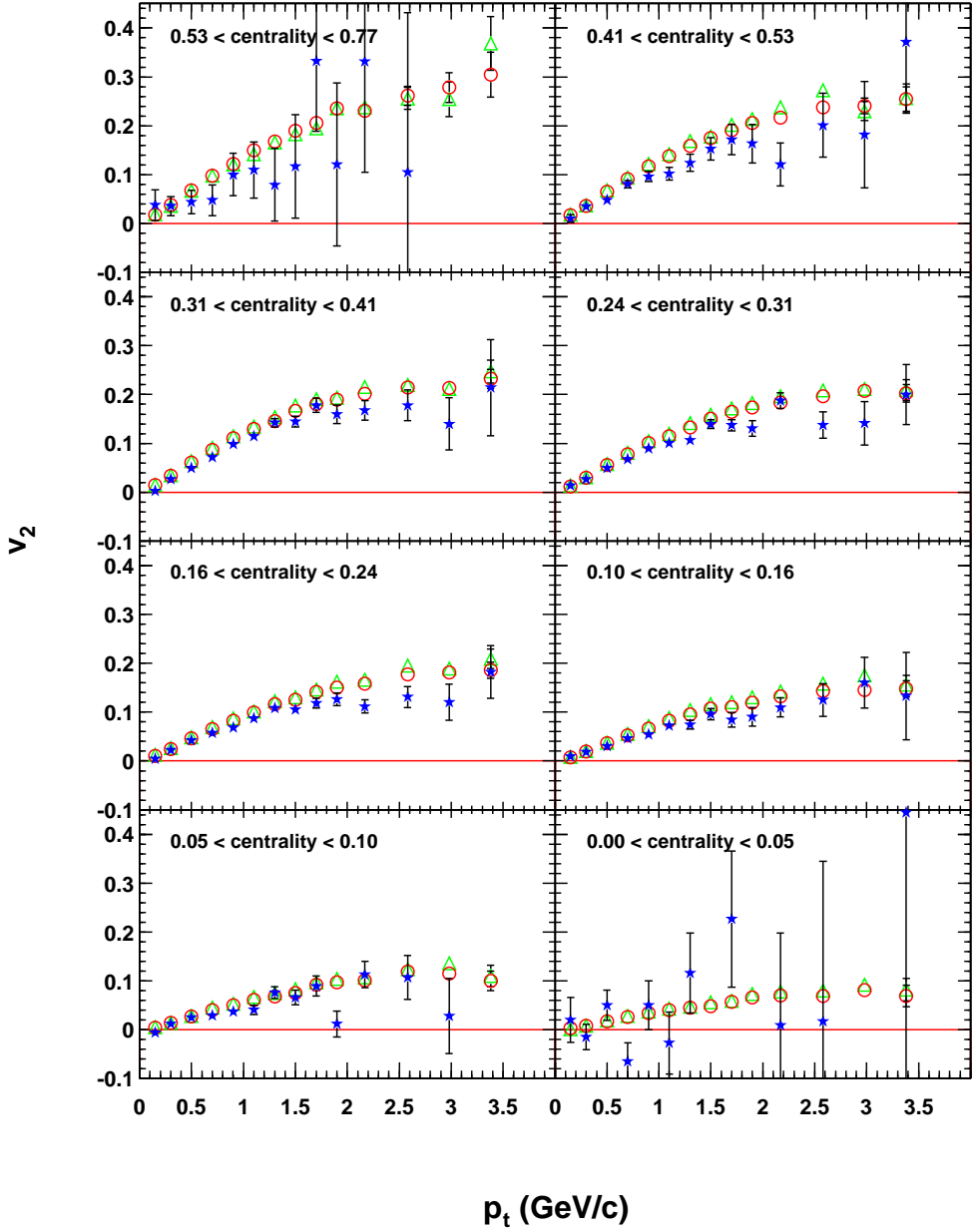}}      
\caption{\label{fig:PtDifferentialFlowAllCent} Reconstructed $v_2$     
versus $p_t$ from the conventional method (circles), from the    
2nd-order cumulant method (triangles), and from the 4th-order cumulant    
method (stars), in eight centrality bins.  The upper left panel shows    
the most peripheral events, and the lower right the most central.}    
\end{figure*}

Figure~\ref{fig:EtaDifferentialFlowAllCent} shows $v_2$ as a function
of pseudorapidity and Fig.~\ref{fig:PtDifferentialFlowAllCent} shows
$v_2$ as a function of transverse momentum.  The eight panels
correspond to the eight bins of relative multiplicity in
Fig.~\ref{fig:vcent} but the centrality is now defined in terms of the
total geometric cross section (see first three columns of
table~\ref{tbl:summaryTable}).  These results illustrate the main
disadvantage of the higher-order cumulant approach compared with any
of the two-particle methods, namely, larger statistical errors, and
this can be seen to be a serious shortcoming in cases where
simultaneous binning in several variables results in small sample
sizes. However, Fig. \ref{fig:vcent} demonstrates that, especially for
the more peripheral bins, the statistical uncertainties for the 
fourth-order cumulant method are smaller than the systematic uncertainties 
for the two-particle methods.

\begin{figure}[ht]      
\resizebox{20pc}{!}{\includegraphics{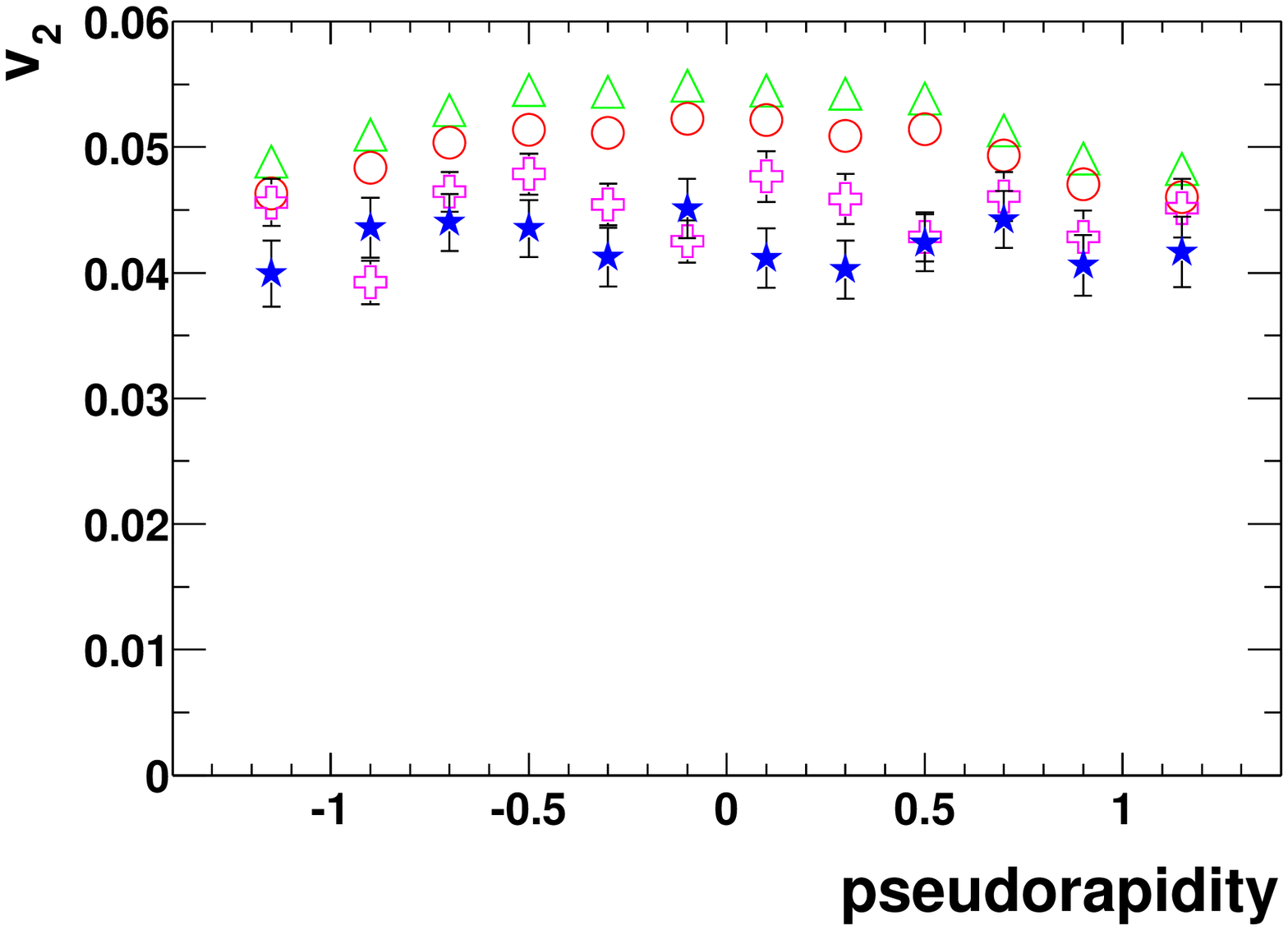}}      
\caption{\label{fig:EtaDifferentialFlowMinBias}        
Elliptic flow versus pseudorapidity from the conventional method    
(circles), from the 2nd-order cumulant method (triangles), from    
quarter-events (crosses), and from the 4th-order cumulant method    
(stars), averaged over all centralities from bin 2 through 7, as    
defined in Figs.~\ref{fig:EtaDifferentialFlowAllCent} and    
\ref{fig:PtDifferentialFlowAllCent}. }    
\end{figure}

\begin{figure}[ht]      
\resizebox{20pc}{!}{\includegraphics{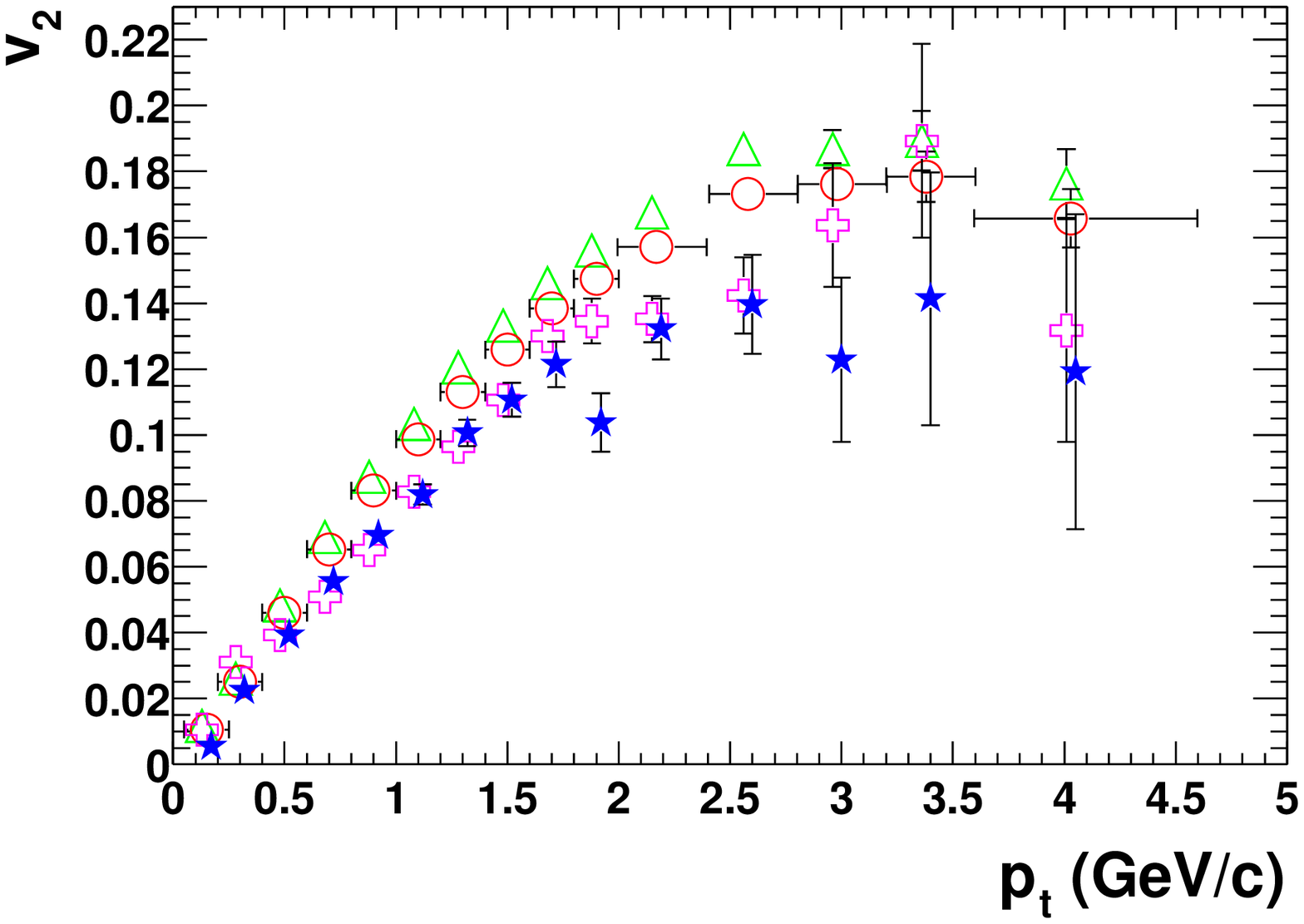}}      
\caption{\label{fig:PtDifferentialFlowMinBias}      
Elliptic flow versus transverse momentum from the conventional method    
(circles), from the 2nd-order cumulant method (triangles), from    
quarter-events (crosses), and from the 4th-order cumulant method    
(stars), averaged over all centralities from bin 2 through 7, as    
defined in Figs.~\ref{fig:EtaDifferentialFlowAllCent} and    
\ref{fig:PtDifferentialFlowAllCent}.}    
\end{figure}

Figures~\ref{fig:EtaDifferentialFlowMinBias} and      
\ref{fig:PtDifferentialFlowMinBias}       
are again plots of elliptic flow versus pseudorapidity and versus
transverse momentum, respectively.  Here the $v_2$ is integrated over
centrality bins 2 through 7.  Bins 1 and 8 are not included in this
average, otherwise they would significantly increase the statistical
error on the result.  The 4th-order cumulant $v_2$ is systematically
about 15\% lower than the conventional pair and cumulant pair
calculations, indicating that non-flow effects contribute to $v_2$
analyses of the latter kind.  The $v_2$ signal based on quarter-events
(as defined in the discussion of Fig.~\ref{fig:vcent}) is closer to
the 4th-order cumulant, although still larger on average, implying
that this pair analysis prescription is effective in removing some,
but not all, non-flow effects.
   
Figure~\ref{fig:PtDifferentialFlowMinBias} verifies that the
$v_2(p_t)$ curve flattens above 2 GeV/$c$~\cite{QM2001Raimond}.  There
is theoretical interest in the question of whether or not $v_2(p_t)$
continues flat at higher $p_t$ or eventually goes down~\cite{GVW01}
--- this issue is the subject of a separate analysis \cite{STARnew},
and the statistics of year-one data from STAR is not suited for
addressing this question via a four-particle cumulant analysis.
            
\begin{figure}[ht]      
\resizebox{20pc}{!}{\includegraphics{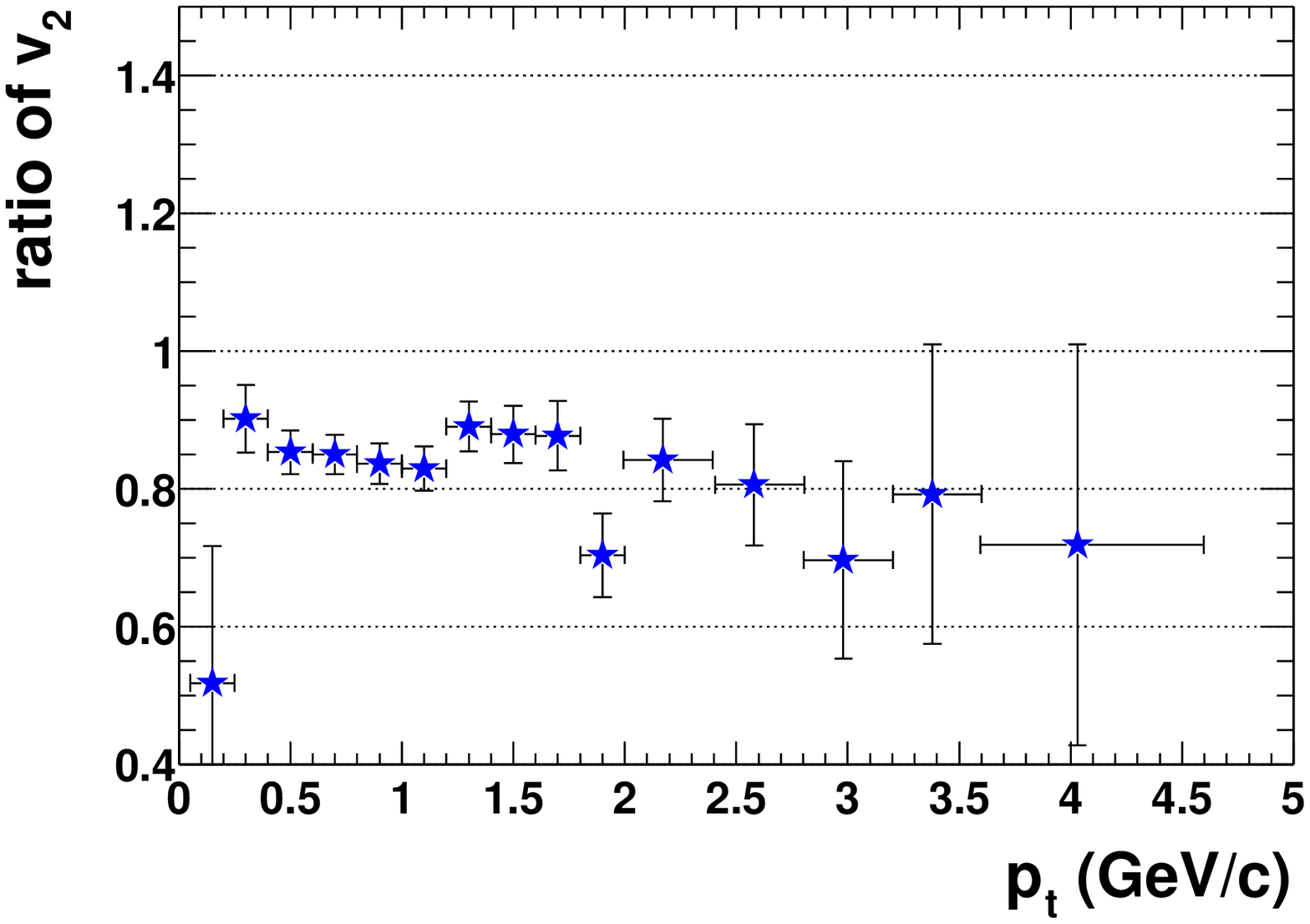}}      
\caption{\label{fig:cumuV2OverStdV2_Pt}      
The ratio of $v_2$ from the 4th-order cumulant divided by $v_2$    
from the conventional method as a function of $p_t$, averaged   
over all centralities from bin 2 through 7, as defined in    
Figs.~\ref{fig:EtaDifferentialFlowAllCent} and    
\ref{fig:PtDifferentialFlowAllCent}.}    
\end{figure}

\begin{figure}[ht]      
\resizebox{20pc}{!}{\includegraphics{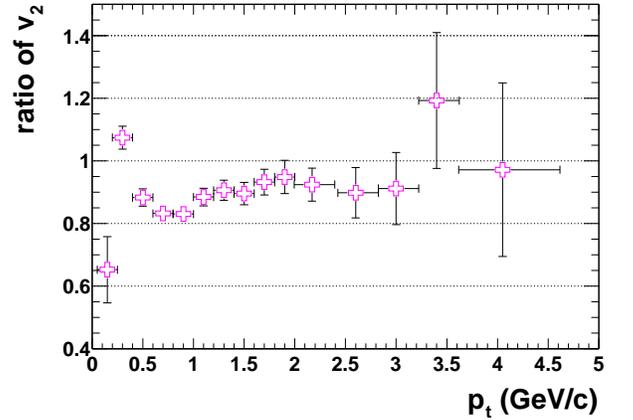}}      
\caption{\label{fig:QtrLowRPV2OverLowRPV2_Pt}      
The ratio of $v_2$ from quarter-events divided by the conventional    
$v_2$ as a function of $p_t$.  In both cases, event planes were    
constructed from low $p_t$ ($< 0.5$ GeV/c) particles. The data are    
averaged over all centralities from bin 2 through 7, as defined in    
Figs.~\ref{fig:EtaDifferentialFlowAllCent} and    
\ref{fig:PtDifferentialFlowAllCent}.}    
\end{figure}

Figure~\ref{fig:cumuV2OverStdV2_Pt} presents the $p_t$-dependence of
the correction factor for non-flow.  Within errors, the relative
non-flow effect is seen to be about the same or increasing very weakly
from low $p_t$ through $p_t \sim 4$ GeV$/c$ --- a somewhat surprising
result, given the presumption that the processes responsible for
non-flow are different at low and high $p_t$.
Fig.~\ref{fig:QtrLowRPV2OverLowRPV2_Pt}, which presents $v_2$ from
quarter-events divided by the conventional $v_2$, both based on event
planes constructed from particles with $p_t < 0.5$ GeV$/c$, offers a
useful insight regarding the approximate $p_t$-independence of
non-flow.  This ratio roughly characterizes the contribution to
non-flow from resonance decays and from other sources which primarily
affect $v_2$ at lower $p_t$, whereas non-flow from (mini)jets ought to
be about equally present in the numerator and the denominator of the
ordinate in Fig.~\ref{fig:QtrLowRPV2OverLowRPV2_Pt}.  A comparison of
Figs.~\ref{fig:cumuV2OverStdV2_Pt} and
\ref{fig:QtrLowRPV2OverLowRPV2_Pt} accordingly does not contradict the
implicit assumption that different phenomena dominate non-flow in
different $p_t$ regions, and implies that the total resultant non-flow
correction by coincidence happens to be roughly the same throughout
the $p_t$ range under study.
   
\begin{figure}[ht]      
\resizebox{20pc}{!}{\includegraphics{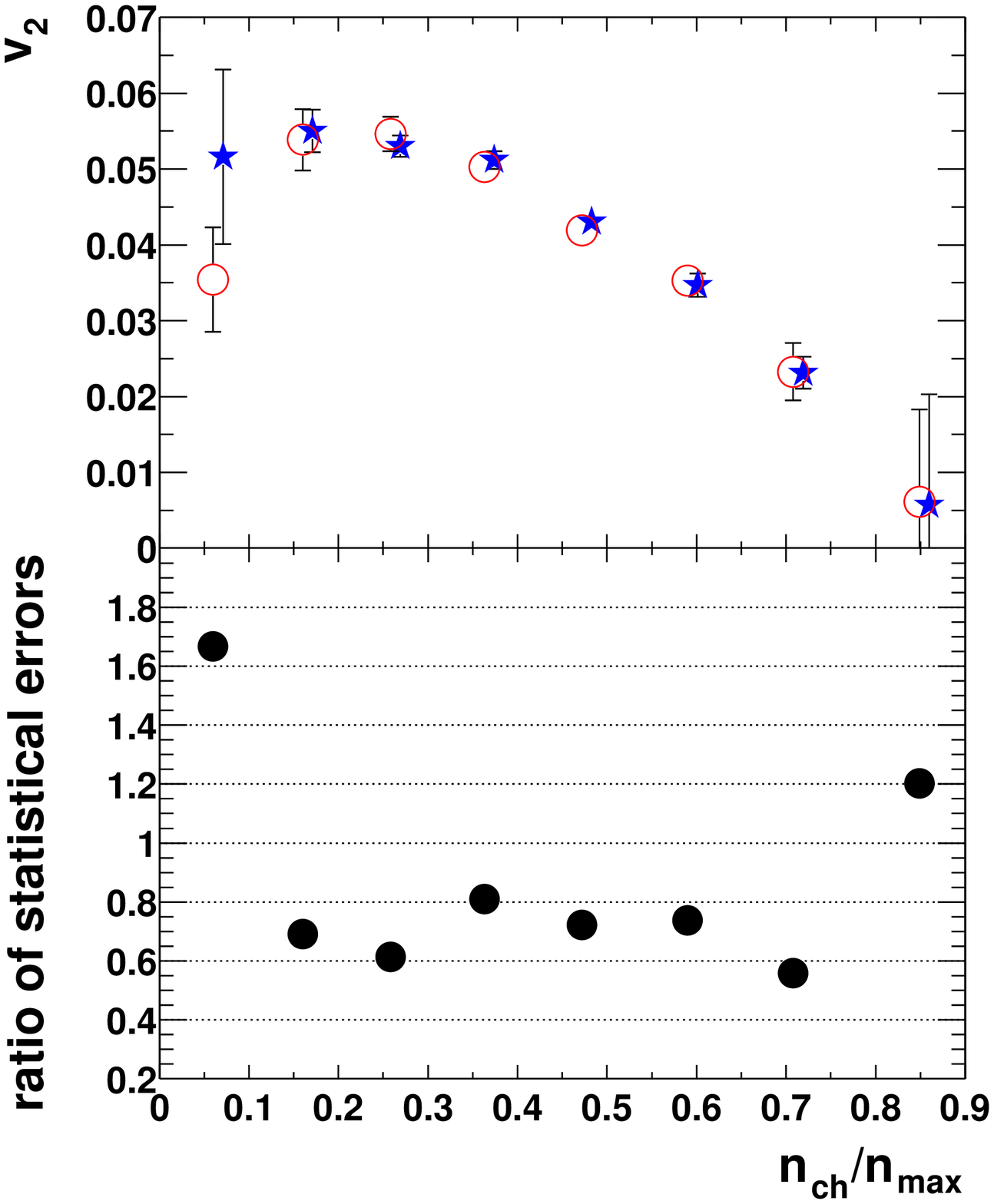}}      
\caption{\label{fig:PtWgtUnitWgt} The upper panel shows measured $v_2$ from 
4th-order cumulants versus centrality with $p_t$ weighting (stars) and unit    
weighting (circles). The bottom panel is the ratio of error from $p_t$
weighted $v_2$ to that of unit weighted $v_2$. }    
\end{figure}

Following the approach of Section II.B, the options of weighting each    
track by either unity or $p_t$ have been compared in the 4th-order    
cumulant analysis.  Fig.~\ref{fig:PtWgtUnitWgt} demonstrates that    
the STAR results are consistent in the two cases, and the $p_t$    
weighting yields smaller statistical errors.  All STAR results    
presented in this paper are computed with $p_t$ weighting unless    
otherwise stated.

\section{Elliptic flow fluctuations}       
         
High precision results presented in this publication become sensitive   
to another effect usually neglected in flow analysis, namely,    
event-by-event flow fluctuations.  The latter can have two   
different origins: ``real'' flow fluctuations --- fluctuations    
at fixed impact parameter and fixed multiplicity (see, for example   
\cite{Kodama}) --- and impact parameter variations among events from    
the same centrality bin in a case where flow does not fluctuate at
fixed impact parameter.  These effects, in principle, are
present in any kind of analysis, including the ``standard'' one based
on pair correlations.  The reason is that any flow measurements are
based on correlations between particles, and these very correlations
are sensitive only to certain moments of the distribution in $v_2$.
In the pair correlation approach with the reaction plane determined
from the second harmonic, the correlations are proportional to $v^2$.
After averaging over many events, one obtains $\la v^2 \ra$, which in
general is not equal to $\la v \ra^2$.  The 4-particle cumulant method
involves the difference between 4-particle correlations and (twice)
the square of the 2-particle correlations.  In this paper,
we assume that this difference comes from correlations in the non-flow
category.  Note, however, that in principle this difference ($\la v^4
\ra - \la v^2 \ra^2 \ne 0$) could be due to flow fluctuations.  Let us
consider an example where the distribution in $v$ is flat from $v=0$
to $v=v_{\rm max}$.  Then, a simple calculation would lead to the
ratio of the flow values from the standard 2-particle correlation
method and 4-particle cumulants as large as $\la v^2 \ra^{1/2}/(2\la
v^2 \ra^2 - \la v^4 \ra)^{1/4} = 5^{1/4}\approx 1.5$.
         
In this study, we consider the possible bias in elliptic flow
measurements under the influence of impact parameter fluctuations
within the studied centrality bins.  The largest effect is expected
within the bin of highest multiplicity, where the impact parameter and
$v_2$ are both known {\it a priori} to fluctuate down to zero in the
limit of the most central collisions.  These fluctuations lead to
bin-width-dependent bias in the extracted $v_2$ measurements.
       
In section III, two approximations were made in order to extract the     
final flow result,        
\begin{displaymath}       
\langle v_n^4 \rangle \simeq \langle v_n^2 \rangle ^2       
~~~~~~ {\rm and} ~~~~~~      
\langle v_n^2 \rangle \simeq \langle v_n \rangle ^2 \,.      
\end{displaymath}       
Taking into account the centrality binning fluctuation on flow, namely
$\sigma_{v_n^2}^2$ and $\sigma_{v_n}^2$,
\begin{displaymath}       
\langle v_n^4 \rangle = \sigma_{v_n^2}^2+ \langle v_n^2 \rangle ^2       
~~~~~~ {\rm and} ~~~~~~      
\langle v_n^2 \rangle = \sigma_{v_n}^2 + \langle v_n \rangle ^2 \,,       
\end{displaymath}       
and Eq.~(\ref{eq:one}) becomes       
\begin{equation}       
-v_n^4 - 2\sigma_{v_n}^2 v_n^2 - \sigma_{v_n}^4 + \sigma_{v_n^2}^2       
= -v_{\rm meas}^4 \,,      
\label{eq:vneqn}       
\end{equation}     
which is a function of $v_n$ and is solvable for $v_n$, if
$\sigma_{v_n}^2$ and $\sigma_{v_n^2}^2$ are known.  A method of
calculating both $\sigma_{v_n}^2$ and $\sigma_{v_n^2}^2$ is now
presented.
         
\begin{figure}[ht]       
\resizebox{20pc}{!}{\includegraphics{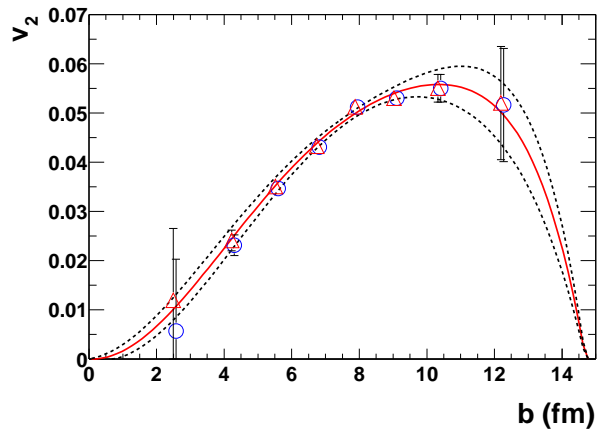}}       
\caption{\label{fig:v2_bCurve} $v_2$ as a function of impact parameter.     
The data points are shown at the values of $\langle b \rangle$ for a 
given centrality bin.  For the description of the fit procedure see text. 
The triangles are the final 4th-order cumulant data after correction
for fluctuations as described in Section IV, while the circles show
the 4th-order cumulant data before this correction.  The dashed lines
represent the estimated uncertainty in the parameterization represented
by the solid curve.}
\end{figure}

First, we need to parameterize $v_n$ as a function of impact
parameter, $b$.  Consider a polynomial fit $v_n = a_0 + a_1 b + ... +
a_6 b^6$, in which case the measured flow is $\langle v_n \rangle =
a_0 + a_1 \langle b \rangle + ... + a_6 \langle b^6 \rangle$.  The
various averages $\langle b \rangle$, $\langle b^2
\rangle$,... $\langle b^{12} \rangle$ can be estimated in each    
centrality bin from filtered HIJING events.  The parameters $a_i$ have    
been determined by minimizing $\chi^2$ in a fit to the eight    
$v_2(n_{\rm ch})$ measurements.  In addition, the fit is constrained    
to go through $v_2 = 0$ at $b = 0$ and at $b_{max}=14.7$    
fm~\cite{Peter2000}. The variation of $b_{max}$ within $\pm 0.5$ fm has 
a negligible effect on $v_2(b)$ at $b < 12$ fm.  
Fig.~\ref{fig:v2_bCurve} shows the resulting curve:   
  
\begin{eqnarray}        
v_2 (b)  = -0.000394\,b +  0.00210\,b^2 \nonumber \\  
-0.0000706\,b^3 -0.0000320\,b^4 +  \nonumber \\ 
0.00000358\,b^5 -1.174\times 10^{-7}\,b^6,  
\label{eq:v2bCurve}       
\end{eqnarray}   
where it is assumed that $b$ is in fm.
In principle, the final corrected $v_2(n_{\rm ch})$ should be    
determined iteratively, but the result is stable on the first    
iteration.    
     
Next we consider      
\begin{eqnarray}       
&& \sigma_{v_n}^2 =     
\langle v_n^2 \rangle - \langle v_n \rangle ^2  = \nonumber \\    
&& (a_0^2 + 2a_0a_1 \langle b \rangle + ... \, a_6^2 \langle b^{12}    
\rangle) -  \nonumber  \\   
&& (a_0 + a_1 \langle b \rangle + ... \, a_6 \langle b^6    
\rangle )^2 \,,     
\end{eqnarray}    
and again the various averages of powers of $b$ can be estimated using    
HIJING.    
      
After computing $\sigma_{v_n}^2$, $\sigma_{v_n^2}^2$, and obtaining
$v_{\rm meas}$ from the four-particle correlation method,
Eq.~(\ref{eq:vneqn}) can be solved to extract the $v_n$ corrected for
impact parameter fluctuations.  The $v_2$ bias is found to be entirely
negligible in all the studied centrality bins except for the most
central, where the correction is about a factor of two (see the
leftmost bin in Fig.~\ref{fig:v2_bCurve}).  In the present analysis,
even a factor of two is not significant due to the large statistical
error on $v_2$ for maximum centrality.  However, the correction to
$v_2$ resulting from finite centrality bin width at maximum centrality
has been determined with lower uncertainty than $v_2$ itself, and will
become important in future studies with large samples of events.
    
Real event-by-event fluctuation in the flow coefficients would also
make the four-particle values lower than the two-particle values.  At
the moment, there is no way to calculate this effect, although it is
expected to be small.

\begin{table*}[hbt]   
\begin{center}   
\begin{tabular}{c|c|c|c|c|c|c}\hline \hline   
  $\langle n_{ch} \rangle$         & $\langle n_{ch}/n_{max} \rangle$  & cross section     &  $ \langle b \rangle$ (fm)  & RMS ($b$) &   $\langle \eps \rangle$     &    $\langle v_2 \rangle$         \\ \hline   
  $53_{33}^{47}$   & $0.060$           & $53-77\%$         &  $12.23 $    & $0.99$   &   $0.420$    &    $0.052 \pm 0.012$ \\   
  $140_{40}^{40}$  & $0.160$           & $41-53\%$         &  $10.36 $    & $0.70$   &   $0.415$    &    $0.055 \pm 0.003$ \\   
  $227_{47}^{43}$  & $ 0.258$          & $31-41\%$         &  $9.06  $    & $0.68$   &   $0.371$    &    $0.053 \pm 0.001$ \\   
  $319_{49}^{41}$  & $0.363$           & $24-31\%$         &  $7.91  $    & $0.64$   &   $0.319$    &    $0.051 \pm 0.001$ \\   
  $415_{55}^{45}$  & $0.472$           & $16-24\%$         &  $6.80  $    & $0.70$   &   $0.261$    &    $0.043 \pm 0.001$ \\   
  $519_{59}^{41}$  & $0.590$           & $10-16\%$         &  $5.56  $    & $0.72$   &   $0.197$    &    $0.035 \pm 0.002$ \\   
  $622_{62}^{38}$  & $0.708$           & $5-10\%$          &  $4.26  $    & $0.80$   &   $0.131$    &    $0.023 \pm 0.002$ \\   
  $746_{86}^{124}$ & $0.849$           & top $5\%$         &  $2.53  $    & $1.00$   &   $0.058$    &    $0.012 \pm 0.015$ \\   
\hline \hline   
\end{tabular}   
\end{center}   
\caption{Tabulated values of observed charged particle multiplicity, 
$n_{ch}/n_{max}$, centrality in terms of percent of total geometric
cross section, impact parameter with spread (root mean square) inferred from 
HIJING, the initial spatial anisotropy $\eps$, and the final corrected 
elliptic flow based on 4th-order cumulants.}
\label{tbl:summaryTable}   
\end{table*}

\section{The centrality dependence of elliptic flow}       
     
\begin{figure}[ht]       
\resizebox{20pc}{!}{\includegraphics{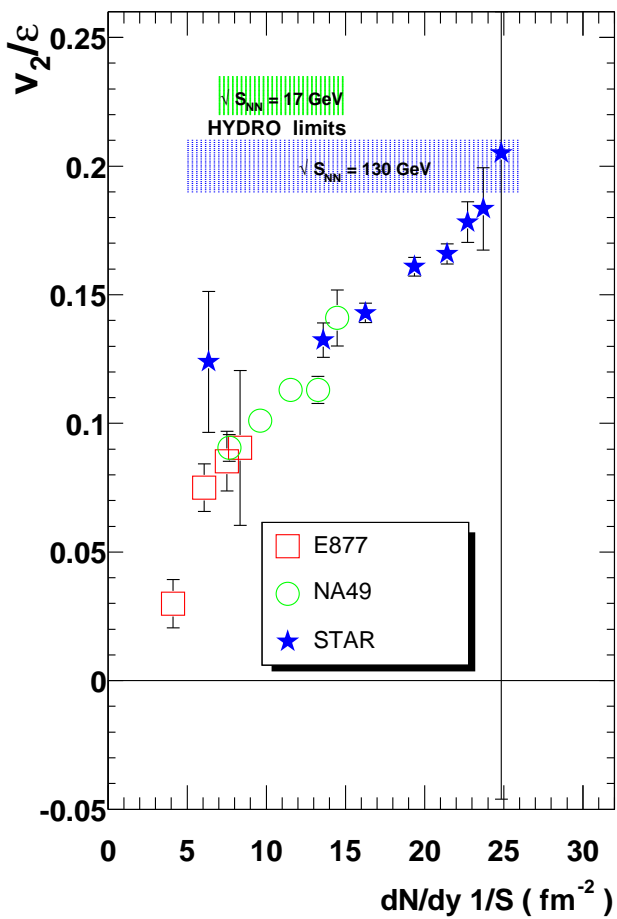}}       
\caption{\label{fig:v2OverEpsilon_density} $v_2/\eps$ as a function   
of charged particle density in Au + Au collisions.  Data are from E877
at the AGS (squares), NA49 at the SPS (circles), and STAR at RHIC
(stars). The AGS and SPS data have been obtained by conventional flow
analysis. The STAR measurements are at $\sqrt{s_{NN}} = 130$ GeV, and
correspond to the final corrected elliptic flow based on 4th-order
cumulants, and we assume $dN/dy = 1.15 dN/d\eta$. The horizontal
shaded bands indicate the hydrodynamic limits for different beam
energies\cite{Kolb00}.  }
\end{figure}

The centrality dependence of elliptic flow is a good indicator of the    
degree of equilibration reached in the reaction \cite{VoloshinCentPhyV2,     
NA49QM}.  Following Ref.~\cite{Peter2000}, we compute the initial spatial   
eccentricity for a Woods-Saxon distribution with a wounded nucleon model   
from     
$$    
\eps = \frac{ \langle y^2 \rangle - \langle x^2 \rangle }      
            { \langle y^2 \rangle + \langle x^2 \rangle }\,    
$$    
where $x$ and $y$ are coordinates in the plane perpendicular to the    
beam and $x$ denotes the in-plane direction. The method of calculation of epsilon is 
the same as used for the hydro values\cite{Kolb00}. The ratio $v_2/\eps$     
is of interest because it has been argued to be independent of centrality     
in a hydrodynamic model with constant speed of sound \cite{Olli92}. 
In hydrodynamic model calculations using an equation of state
with a phase transition (sound speed is not constant) this ratio does 
change as a function of centrality, however within the $10\%$ level \cite{Kolb00}.    
Hydrodynamics represents one possible limiting case in describing nuclear    
collisions --- the limit where the mean free path for interaction of the    
constituents represented by the fluid cells is very small compared with    
the region of nuclear overlap.  The opposite limit, where the mean free    
path is long (or at least comparable to the dimensions of the nuclear    
overlap region) is normally known as the Low Density Limit (LDL).  In    
nuclear transport models, the mean number of hard binary interactions per    
particle is typically small, and the predictions of these models tend to be    
closer to the low density limit than the hydro limit.  In order to judge     
the proximity of measured flow data to either of these limits, it is    
useful to plot, as in Fig.~\ref{fig:v2OverEpsilon_density}, $v_2/\eps$ versus    
charged particle density in the form $(dN/dy)/S$, where $dN/dy$ is rapidity    
density, and the area of the overlap region is $S = \pi \sqrt{ \la x^2 \ra    
\la y^2 \ra }$ as computed above.  Since $v_2/\eps$ is proportional to    
$(dN/dy)/S$ in the LDL case \cite{Heiselberg,VoloshinCentPhyV2}, this    
form of plot offers meaningful insights without reference to detailed    
theoretical models.     
   
Figure~\ref{fig:v2OverEpsilon_density} presents Au + Au data from AGS/E877   
\cite{Barr95}, from NA49 \cite{NA49QM}, as well as the current    
STAR measurements based on 4th-order cumulants, corrected for
fluctuations as detailed in Section IV.  Alternative forms of the
centrality dependence readily can be generated using the tabulated
quantities presented in Table I.  Generally, the current STAR results
underline the need for much increased statistics, particularly for the
most central collisions.  Within the uncertainties, a smooth trend of
increasing $v_2/\eps$ with increasing centrality (larger $(dN/dy)/S$)
is observed, without the obvious kink that has been suggested as a phase
transition signature~\cite{Sorge99, Heiselberg}.  Another proposed
phase transition signature which is not favored by the data is a few
percent rise in $v_2/\eps$ with {\it decreasing} centrality
\cite{Kolb00}.  It is noteworthy that the $v_2/\eps$ values reached in
the most central RHIC collisions are consistent with the hydrodynamic
limit \cite{Olli92, Kolb99, Kolb00}, whereas $v_2/\eps$ in central
collisions at AGS and SPS is significantly lower. It is also worthy of
note that while the roughly linear relationship between $v_2/\eps$ and
$(dN/dy)/S$ across the presented beam energies and centralities is
consistent with the LDL picture~\cite{Heiselberg}, the measured
$v_2(p_t)$ Fig.~\ref{fig:PtDifferentialFlowAllCent} cannot be
explained by current LDL implementations \cite{Kolb01}, and is much
closer to hydrodynamic calculations up to 2 GeV$/c$ \cite{Kolb01}.

\section{Conclusion}       
       
In this work, we provide details of the approach for treating non-flow
correlations within the framework of the standard elliptic flow
analysis method based on particle pairs.  We also compare the standard
method with a new and simpler pair analysis based on the scalar
product of flow vectors.  The latter yields a 15 -- 35\% reduction in
statistical errors, with the best improvement occurring in the case of
the most central and the most peripheral events.
    
It is concluded that four-particle correlation analyses can reliably
separate flow and non-flow correlation signals, and the latter account
for about 15\% of the observed second-harmonic azimuthal correlation
in year-one STAR data.  The cumulant approach has demonstrated some
advantages over the previous alternatives for treating non-flow
effects.  In particular, 4th-order cumulants allows us to present
$v_2$ measurements fully corrected for non-flow effects, in contrast
to the earlier analyses where the non-flow contribution was partly
removed and partly quantified by the reported systematic
uncertainties.  It is observed that non-flow correlations are present
in $\sqrt{s_{NN}} = 130$ GeV Au + Au events throughout the studied
region $|\eta| < 1.3$ and $0.1 < p_t < 4.0$ GeV/$c$, and are present
at all centralities.  The largest contribution from non-flow
correlations is found among the most peripheral and the most central
collisions.
         
On the other hand, a 4th-order cumulant analysis is subject to larger    
statistical errors than a conventional pair correlation analysis of    
the same data set.  The total uncertainty on the 4th-order 
analysis, including both statistical and systematic effects, is smaller 
for year-one STAR data except in the most central and peripheral panels of 
Figs. \ref{fig:EtaDifferentialFlowAllCent} and 
\ref{fig:PtDifferentialFlowAllCent}. In the case of future studies of     
larger numbers of events, a higher-order analysis should provide an     
advantage in all cases.    
          
Fluctuations within the studied multiplicity bins have the potential
to bias elliptic flow results.  This bias has been estimated and found
to be entirely negligible except for the most central multiplicity
bin, where the correction is about a factor of two.  In the present
analysis, even this large a bias is only marginally significant, but
again, this correction will presumably be important in future
studies with much improved statistics.
    
We present STAR data for $v_2/\eps$ --- elliptic flow in various
centrality bins, divided by the initial spatial eccentricity for those
centralities.  Mapping centrality onto a scale of charged particle
density enables us to study a broad range of this quantity, from
peripheral AGS collisions, through SPS, and ending with central RHIC
collisions.  Within errors, the STAR data follow a smooth trend. No
evidence for a softening of the equation of state or for a change in
degrees of freedom has been observed.  The three experiments at widely
differing beam energies show good agreement in $v_2/\eps$ where they
overlap in their coverage of particle density.  The pattern of
$v_2/\eps$ being roughly proportional to particle density continues
over the density range explored at RHIC, which is consistent with a
general category of models which approximate the low density limit as
opposed to the hydrodynamic limit.  Nevertheless, $v_2/\eps$ at
STAR is consistent with having just reached the hydrodynamic limit for
the most central collisions.

\begin{acknowledgments}       
We thank Nicolas Borghini, Jean-Yves Ollitrault, and Mai Dinh for    
helpful discussions and suggestions.   
We wish to thank the RHIC Operations Group and the RHIC Computing Facility
at Brookhaven National Laboratory, and the National Energy Research 
Scientific Computing Center at Lawrence Berkeley National Laboratory
for their support. This work was supported by the Division of Nuclear 
Physics and the Division of High Energy Physics of the Office of Science of 
the U.S. Department of Energy, the United States National Science Foundation,
the Bundesministerium fuer Bildung und Forschung of Germany,
the Institut National de la Physique Nucleaire et de la Physique 
des Particules of France, the United Kingdom Engineering and Physical 
Sciences Research Council, Fundacao de Amparo a Pesquisa do Estado de Sao 
Paulo, Brazil, the Russian Ministry of Science and Technology and the
Ministry of Education of China and the National Natural Science Foundation 
of China.
\end{acknowledgments}


\begin{thebibliography}{99}       
       
\bibitem{Reis97}       
       
  W. Reisdorf, and H. G. Ritter,        
  Annu. Rev. Nucl. Part. Sci. {\bf 47}, 663 (1997).       
       
\bibitem{Herr99}       
       
  N. Herrmann, J. P. Wessels, and T. Wienold,       
  Annu. Rev. Nucl. Part. Sci. {\bf 49}, 581 (1999).       
          
\bibitem{OlliQM98}    
    
  J.-Y. Ollitrault, Nucl. Phys. {\bf A638}, 195c (1998).    
    
\bibitem{Palaiseau}    
    
  A. M. Poskanzer, e-print nucl-ex/0110013 (2001).    
    
\bibitem{Sorge97}    
    
  H. Sorge,     
  Phys. Rev. Lett. {\bf 78}, 2309 (1997).    
    
\bibitem{Olli92}       
       
  J.-Y. Ollitrault,        
  Phys. Rev.  {\bf D46}, 229 (1992).       

       
\bibitem{Volo96}       
       
  S. Voloshin and Y. Zhang,        
  Z. Phys.  {\bf C70}, 665 (1996).       
         
\bibitem{Posk98}       
       
  A. M. Poskanzer and S. A. Voloshin,       
  Phys. Rev.  {\bf C58}, 1671 (1998).       
       
\bibitem{STAR01}       
       
  STAR collaboration,  K. H. Ackermann {\it et al.},       
  Phys. Rev. Lett. {\bf 86}, 402 (2001).       
          

\bibitem{Teaney02}

  D. Teaney, J. Lauret, and Edward V. Shuryak,
  Nucl. Phys. {\bf A698}, 479 (2002).

\bibitem{Teaney01}
  D. Teaney, J. Lauret, and Edward.V. Shuryak,
  e-print nucl-th/0110037 

\bibitem{ZiWei02}

  Zi-wei Lin and C. M. Ko, 
  Phys. Rev.  {\bf C65}, 034904 (2002).       
  

\bibitem{Ko02}

  C. M. Ko, Zi-wei Lin, and S. Pal,
   e-print nucl-th/0205056 (2002). 

\bibitem{Molnar}

  D. Molnar and M. Gyulassy,
  Nucl. Phys. {\bf A697}, 495 (2002).

\bibitem{Zabrodin}

  E.E. Zabrodin, C. Fuchs, L.V. Bravina, and A. Faessler,
  Phys. Lett. {\bf B508}, 184 (2001).
  

\bibitem{Huma02}

  T. J. Humanic, 
  e-print nucl-th/0205053. 

\bibitem{STAR01b}       
       
  STAR collaboration,  C. Adler {\it et al.},       
  Phys. Rev. Lett. {\bf 87}, 182301 (2001).      

\bibitem{Heiselberg}    
    
  H. Heiselberg and A.-M. Levy,     
  Phys. Rev.  {\bf C59}, 2716 (1999).    
 
\bibitem{Jian92}       
       
  J. Jiang {\it et al.},       
  Phys. Rev. Lett. {\bf 68}, 2739 (1992).       
 
         
\bibitem{Olli00}       
       
  N. Borghini, P. M. Dinh, and J.-Y. Ollitrault,         
  Phys. Rev.  {\bf C63}, 054906 (2001).       
   
\bibitem{Olli01}     
    
  N. Borghini, P. M. Dinh, and J.-Y. Ollitrault,         
  Phys. Rev.  {\bf C64}, 054901 (2001).       
       
\bibitem{STAR99}       
       
  STAR collaboration,  K. H. Ackermann {\it et al.},       
  Nucl. Phys. {\bf A661}, 681c (1999).       
     
\bibitem{Dani85}       
       
  P. Danielewicz and G. Odyniec,       
  Phys. Lett. {\bf B157}, 146 (1985).         
       
\bibitem{HIJING}    
    
  M. Gyulassy and X.-N. Wang, Comput. Phys. Commun. {\bf 83},    
  307 (1994);    
  X.N. Wang and M. Gyulassy, Phys. Rev.  {\bf D44}, 3501 (1991).    
       
\bibitem{Dani95}       
       
  P. Danielewicz,       
  Phys. Rev.  {\bf C51}, 716 (1995).       
 
\bibitem{ArtSergeiLBLReport}       
         
  A. M. Poskanzer and S. A. Voloshin,       
  {\it LBNL Annual Report}       
  http://ie.lbl.gov/nsd1999/rnc/RNC.htm R34 (1998).       
  
\bibitem{E877PRL94}

  E877 collaboration, J. Barrette {\it et al.},  
            Phys. Rev. Lett.  {\bf 73}, 2532 (1994);  

\bibitem{Olli95}

  J.-Y. Ollitrault, Nucl. Phys. {\bf A590}, 561c (1995).

\bibitem{Olli97}

  J.-Y. Ollitrault, 
   e-print nucl-ex/9711003 (1997). 

\bibitem{borghiniNoflow}       
       
  N. Borghini, P. M. Dinh, and J.-Y. Ollitrault,         
  Phys. Rev.  {\bf C62}, 034902 (2000).       
       
\bibitem{Mai}       
       
  P. M. Dinh, N. Borghini, and J.-Y. Ollitrault,         
  Phys. Lett.  {\bf B477}, 51 (2000).       
    
\bibitem{Biya81}       
    
  M. Biyajima, Phys. Lett. {\bf B92}, 193 (1980);      
  Prog. Theor. Phys. {\bf 66}, 1378 (1981).       
       
\bibitem{Libo89}       
       
  R. L. Liboff,        
  {\it Kinetic Theory} Prentice Hall, Englewood Cliffs, New Jersey       
  (1989).       
          
\bibitem{Egge93}       
       
  H. C. Eggers, P. Lipa,  P. Carruthers, and  B. Buschbeck,       
  Phys. Rev.  {\bf D48}, 2040 (1993).       
          
\bibitem{cumuPraticeGuide}     
    
  N. Borghini, P. M. Dinh, and J.-Y. Ollitrault,         
  e-print nucl-ex/0110016.     
       
\bibitem{mevsim}    
    
  R. L. Ray and R. S. Longacre,     
  STAR Note SN0419, 1999, e-print nucl-ex/0008009.    
        
\bibitem{QM2001Raimond}       
       
  STAR collaboration, R. Snellings {\it et al.},     
  Quark Matter, (2001).       
       
\bibitem{GVW01}    
    
  M. Gyulassy, I. Vitev, and X. N. Wang,     
  Phy. Rev. Lett. {\bf 86}, 2537 (2001).    
    
\bibitem{STARnew}       
       
  STAR collaboration,  C. Adler {\it et al.},  
  in preparation.      
       
\bibitem{Kodama}

  C.E. Aguiar {\it et al.},
  Nucl. Phys.  {\bf  A698}, 639c (2002).

\bibitem{Peter2000}       
       
  P. Jacobs and G. Cooper,        
  STAR Note SN0402, 1999, e-print nucl-ex/0008015.   

\bibitem{VoloshinCentPhyV2}       
       
  S. A. Voloshin and A. M. Poskanzer,       
  Phys. Lett.  {\bf B474}, 27 (2000).    
       
\bibitem{NA49QM}    
    
  A. M. Poskanzer and S. A. Voloshin,    
  Nucl. Phys.  {\bf A661}, 341c (1999).    
   
\bibitem{Kolb00}       
       
  P. F. Kolb, J. Sollfrank, and U. Heinz,       
  Phys. Rev.  {\bf C62}, 054909 (2000).    
    

\bibitem{Barr95}     
  
  E877 collaboration, J. Barrette {\it et al.},  
          Phys. Rev.  {\bf C51}, 3309 (1995);  
  {\bf C55}, 1420 (1997).    
      

\bibitem{Sorge99}    
    
  H. Sorge, Phys. Rev. Lett. {\bf 82}, 2048 (1999).    
    
   
\bibitem{Kolb99}       
       
  P. Kolb, J. Sollfrank, and U. Heinz,       
  Phys. Lett.  {\bf B459}, 667 (1999).    
    
\bibitem{Kolb01}    
    
  P.F. Kolb, P. Huovinen, U. Heinz, and H. Heiselberg,
  Phys. Lett.  {\bf B500}, 232 (2001).          
   
\end{thebibliography}
\end{document}